\documentclass[twocolumn,showpacs,preprintnumbers,amssymb]{revtex4}
\usepackage{pdfpages}
\usepackage{graphics}
\usepackage{epsfig} 

\usepackage{graphics}
\usepackage{color}      % use if color is used in text
\usepackage{graphicx}
\usepackage{graphicx}% Include figure files
\usepackage{dcolumn}% Align table columns on decimal point

\usepackage{float}  

\usepackage{amsmath,amssymb}
\usepackage{graphicx}

\begin{document}

\title{ Stochastic Modeling of HIV Reactivation Under ART Washout and Immune Fluctuations}

\author{Mesfin Asfaw  Taye}
\affiliation {West Los Angeles College, Science Division \\9000  Overland Ave, Culver City, CA 90230, USA}%Lines break automatically or can be forced with \\

\email{tayem@wlac.edu}

\begin{abstract}
Post-treatment HIV-1 rebound is dictated  by the stochastic reactivation of latent reservoirs influenced by immune fluctuations and antiretroviral therapy (ART) decay. In this paper, we study the effects of time, immune variability, and ART pharmacokinetics on HIV reactivation following treatment interruption. In this work, we model   latency reversal as a Poisson-driven stochastic process shaped by circadian rhythms, transient inflammation, and immune bursts. We incorporate gamma-distributed waiting times in order to capture heterogeneity in activation dynamics and confirm the reduced early post-ART reactivation. Building on earlier stochastic models of reactivation,  we present a rigorous framework that captures a range of activation dynamics—including constant, sinusoidal, stochastic, and exponentially decaying rates—reflecting both immune-driven variability and pharmacokinetic influences on latency reversal. We further investigate block-and-lock strategies, showing that combining ART with latency-promoting agents pharmacologically stabilizes the reservoir and delays reactivation, particularly when drug decay is slow.  In addition, we study a deterministic model describing the dynamics of viruses, latent cells, and infected cells, including the role of the immune response and the effect of shock-and-kill strategies. Our results reveal that synchronizing latency-reversing agents with immune activation cycles enhances viral clearance and delays rebound. This framework refines post-treatment control predictions and generalizes to persistent infections, such as hepatitis B virus (HBV) and cytomegalovirus (CMV), where immune fluctuations and drug decay shape stochastic viral reactivation.
\end{abstract}

\pacs{Valid PACS appear here}% PACS, the Physics and Astronomy
                             % Classification Scheme.
%\keywords{Suggested keywords}%Use showkeys class option if keyword
                              %display desired
\maketitle

\section{Introduction}

The introduction of antiretroviral therapy (ART) has led to the improvement  of   HIV-1 management since the drug  suppresses  viral replication and delays  disease progression \cite{Murray2016, Whitney2014, Okoye2018}. However, a complete   cure remains impossible due to long-lived, latently infected CD4$^+$ T cells  that  are established early in infection \cite{Chun1997, Siliciano2003, Li2020}. These reservoirs persist via transcriptional silencing and immune evasion \cite{Wu2020, Pinkevych2019, Fennessey2017}. As a result, when  treatment  is interrupted,   stochastic reactivation and rapid viral rebound occur  \cite{Byrareddy2016, Whitney2018, Hill2018}. Mathematical modeling has been pivotal for elucidating viral and latency dynamics \cite{mes2, mes1, mes3, mes4, mu1, mu2, mu3, mu4, mu5, mu6, mu7, mu8, mu9, mu10, mu11, mu12, mu14, Hill2014, Pinkevych2016, DeScheerder2019}.  However, most of these  early deterministic approaches overlook the heterogeneity of reactivation, and it has recently been confirmed that   latency dynamics  are shaped by immune perturbations and environmental factors \cite{Lorenzo2016, Christensen-Quick2018, Zens2014, Bukrinsky1991}. This warrants  stochastic frameworks that incorporate  reactivation variability, immune fluctuations, and ART pharmacokinetics to obtain improved rebound prediction \cite{Ribeiro2022}. Several stochastic models and mathematical frameworks have recently been developed to analyze complex biological and networked systems, including hepatitis B dynamics with media effects, co-infection modeling, Lévy-driven transmission, stochastic control strategies, delay-based vaccination models, and fractional-order epidemic systems \cite{hbv1,hbv2,hbv3,hbv4,hbv5,worm1,worm2,burke2022}.

Several empirical studies and non-human primate (NHP) trials indicate  considerable  heterogeneity in viral rebound timing following ART cessation \cite{Wu2020, Pinkevych2019}  due to  the interplay of immune thresholds, epigenetic regulation, and ART washout kinetics \cite{Fennessey2017}. These research works confrim   that   early reactivation events are infrequent, with later rebounds shaped by circadian rhythms, inflammatory episodes, and immune activation bursts \cite{Christensen-Quick2018, Zens2014}.  As these  complexities are poorly captured by standard ordinary differential equation (ODE) models,  stochastic differential equation (SDE) approaches that account for fluctuating reactivation rates and immune-mediated clearance \cite{Hill2018, Davenport2019} are required.  Frequency-dependent models further include  variability in latency reversal, immune surveillance, and ART pharmacodynamics \cite{Pinkevych2016, DeScheerder2019}. Laboratory and in vivo studies have significantly advanced our understanding of HIV latency. Moreover Jurkat-derived T cell lines have shown  mechanisms of chromatin-based silencing, while primary CD4$^+$ T cell models better capture the heterogeneity of latent reservoirs \cite{Lewis2015, Bosque2009}. In non-human primates (NHPs), latency-reversing agents (LRAs) and immune-based therapies have been evaluated \cite{Marsden2018}, with combination approaches, such as Toll-like receptor (TLR) agonists paired with broadly neutralizing antibodies (bNAbs) and  this shows  greater efficacy in delaying viral rebound than monotherapies \cite{Borducchi2018}.

Despite progress in targeting HIV latency, still several  key challenges remain. Single-agent latency-reversing agents (LRAs) have shown limited efficacy, which highlights  the need for combination approaches that pair reactivation with immune clearance \cite{Archin2012, Borducchi2018}. The stochastic nature of reactivation suggests that interventions should align with periods of heightened immune responsiveness \cite{Wu2020, Byrareddy2016}. Moreover, reservoir heterogeneity, immune variability, and unpredictable latency dynamics continue to hinder sustained ART-free remission \cite{Christensen-Quick2018, Bukrinsky1991, van}  and this calls  for  better  strategies that integrate immunological targeting with individualized treatment dynamics.

In this work, we develop both stochastic and deterministic models in order to study   HIV reactivation dynamics following  ART interruption. The stochastic framework includes  time-dependent activation rates, gamma-distributed latency times, and pharmacokinetic drug decay in order to capture   the effects of circadian rhythms, immune fluctuations, and heterogeneous reactivation. We analyze multiple activation profiles  such as  constant, sinusoidal, stochastic, and decaying rates. We then derive analytical expressions for cumulative activation, survival probability, and viral load. We further investigate block-and-lock strategies, showing that combining ART with latency-promoting agents pharmacologically stabilizes the reservoir and delays reactivation, particularly when drug decay is slow.  Complementing this, our deterministic model captures the coupled dynamics of latent, infected, and target cells under time-varying ART and LRA exposures. Overall,  these models show that  immune-driven fluctuations lead to clustered reactivation events and  slow ART washout delays rebound. Optimal post-treatment control  can be  achieved when latency-reversing agents are administered during immune activation peaks and  ART clearance is gradual. This unified framework provides mechanistic insights and quantitative guidance for optimizing the timing and efficacy of shock-and-kill interventions.

The remainder of this paper is structured  as follows. In Section II,  we discuss the Poisson process of HIV reactivation. In Section III, we consider  HIV reactivation with a gamma-distributed  waiting time. In Section IV, we present a mathematical model of viral reactivation. We study the  dynamics  of  the virus, latent cells, and  infected cells.  In Section V, we present the shock and kill mathematical model.  Section VI presents the summary and conclusions.

\section{Poisson Process for HIV Reactivation with Drug Washout }

Based on the work  Wu et al.~\cite{pink}, we develop a rigorous time-dependent Poisson model to study HIV reactivation under four activation profiles: constant, sinusoidal, stochastic, and exponentially decreasing during ART washout. The exact solutions for cumulative activation and inter-event times reveal how rate variability drives event clustering and rebound dynamics. We extend this with an SDE framework integrating immune fluctuations and ART pharmacokinetics. This  captures how drug decay and stochastic perturbations shape latency reversal.

\subsection{Stochastic Reactivation with Constant Activation Rate}

A fundamental challenge in understanding post-treatment HIV reactivation comes from  the stochastic nature of latency reversal and viral rebound following antiretroviral therapy (ART) cessation. In this work,  by developing  a Poisson-driven stochastic framework that accounts for immune perturbations, ART pharmacokinetics, and latency-reversing agent (LRA), we address this complexity.  The probability $P(n,t)$ of observing $n$ reactivated cells at time $t$  is given aby
\begin{equation} 
\frac{dP(n,t)}{dt} = \lambda \left[ (n-1) P(n-1,t) - n P(n,t) \right]. 
\end{equation}
This equation  governs transitions between activation states, ensuring that all probability masses remain at $n = 0$ prior to ART discontinuation. At the moment of drug washout ($t = t_w$), no reactivation events occurs
$
P(n,t_w) = \delta_{n,0}
$.
Here, $\delta_{n,0}$ denotes the Kronecker delta. After washout, reactivation follows a Poisson process with rate $\lambda$, yielding a shifted counting process:
$
N_{t - t_w} \sim \text{Poisson}(\lambda (t - t_w))
$
We confirm that the expected number of reactivation events is as follows:
\begin{eqnarray} 
E[N_t] &=& 0, \quad t < t_w, \\ \nonumber 
       &=& \lambda (t - t_w), \quad t \geq t_w,
\end{eqnarray}
demonstrating a linear increase post-washout.  We define \(t_w\) as the ART washout threshold: for \(t < t_w\), ART levels remain high and reactivation is effectively blocked (\(\lambda(t) \approx 0\)); for \(t \geq t_w\), drug concentrations decay and reactivation becomes possible.

Our findings also  align with those of Van Dorp et al. \cite{van}. Explicitly solving their stochastic multi-reactivation model, we recover their key result describing viral  growth
\begin{equation}  
V_t = v_0 \sum_{i=1}^{N_t} e^{g (t - T_i)},
\end{equation}  
where each reactivated latent cell contributes exponentially to viral replication at rate $g$. Here, \( T_i \) denotes the activation times, which are random variables drawn from a nonhomogeneous Poisson process. The expected viral load post-ART interruption is as follows:
\begin{eqnarray}  
E[V_t] &=& 0, \quad t < t_w, \\ \nonumber  
       &=& \frac{v_0 \lambda}{g} (e^{g(t - t_w)} - 1), \quad t \geq t_w,  
\end{eqnarray}  
confirming an exponential increase driven by the reactivation rate $\lambda$ and replication dynamics. The variance that exhibits significant stochastic fluctuations
\begin{eqnarray}  
\text{Var}[V_t] &=&0, \quad t < t_w, \\ \nonumber  
                 &=& \frac{v_0^2 \lambda}{2g} (e^{2g (t - t_w)} - 1), \quad t \geq t_w,  
\end{eqnarray}  
highlights  the limitations of deterministic models in capturing latency-reversal dynamics. A critical question in treatment interruption studies is the probability of a delayed viral rebound. The survival probability $P(T > t)$ follows an exponential decay:
\begin{eqnarray} 
P(T > t) &=& 1, \quad t < t_w, \\ \nonumber 
         &=& e^{-\lambda (t - t_w)}, \quad t \geq t_w, 
\end{eqnarray}
This implies that the likelihood of reactivation increases over time. The waiting time to the first activation follows an exponential distribution, with the expected time:
\begin{equation} 
E[T_1] = t_w + \frac{1}{\lambda}, 
\end{equation}
Experimental estimates suggest reactivation rates between $0.17$ and $0.54$ per day for HIV and $0.5$ and $2.1$ per day for SIV.  The expected waiting times rang from 1.9 to 6 days post-washout \cite{Wu2020}.  More generally, the $k$-th activation event follows a gamma-distributed waiting time:
$
T_k \sim \text{Gamma}(k, \lambda),
$
with expected value:
\begin{equation} 
E[T_k] = t_w + \frac{k}{\lambda}, 
\end{equation}
confirming that the expected time to the $k$th activation increases linearly with the number of activation events. By maintaining reservoir heterogeneity and stochastic fluctuations in latency reversal, this  gamma-distributed waiting time framework provides a statistical foundation for modeling multiple reactivation events.

 Finally, for a constant activation rate $\lambda_0$, the survival probability  is given as 
\begin{equation} 
S(t - t_w) = \exp(-\lambda_0 (t - t_w)),
\end{equation}
This equation  is vital in  describing the probability of remaining in latency without reactivation, and it also  shows the fundamental timescale for viral rebound under continuous stochastic activation. This also helps gain a basic  understanding of HIV latency reversal and treatment interruption strategies by  providing a rigorous probabilistic framework for optimizing post-ART control and latency-reversing interventions.

\subsection{Sinusoidally Modulated Activation Rate}

Let us now consider a time-dependent Poisson process with a sinusoidally varying activation rate 
\begin{equation}
\lambda(t) = \lambda_0 + A \sin(\omega t),
\end{equation}
where $\lambda_0$, $A$  and  $\omega$  denote the baseline activation rate, amplitude of oscillatory modulation, and  denotes the frequency of periodic fluctuations, respectively. Since  activation events follow a cyclic pattern, such as circadian rhythms, neuronal firing, and viral reactivation,  this   formulation is particularly relevant for biological and physical systems. 

The cumulative rate function  that quantifies  the total expected activations up to time $t$, is given by
\begin{equation}
\Lambda(t) = \int_{t_w}^{t} \lambda(s) ds.
\end{equation}
We rewrite the  above equation  as 
\begin{equation}
\Lambda(t) = \lambda_0 (t - t_w) - \frac{A}{\omega} \cos(\omega t) + \frac{A}{\omega} \cos(\omega t_w).
\end{equation}

Similarly we write  the expected number of activations  as 
\begin{equation}
E[N_t] = \Lambda(t) = \lambda_0 (t - t_w) - \frac{A}{\omega} \cos(\omega t) + \frac{A}{\omega} \cos(\omega t_w),
\end{equation}
The equation  reveals the impact of periodic fluctuations  and it shows that   activation rates oscillate around their baseline level, contributing to the structured clustering of events.

Since inter-event times are inversely proportional to the instantaneous rate, the waiting time between consecutive activations is given by 
\begin{equation}
E[T_{n+1} - T_n] = \frac{1}{\lambda(t)},
\end{equation}
During peak activation phases ($\sin(\omega t) \approx 1$), the  waiting time has a form
\begin{equation}
E[T_{n+1} - T_n] \approx \frac{1}{\lambda_0 + A}.
\end{equation}
For large oscillatory amplitudes, we get 
\begin{equation}
E[T_{n+1} - T_n] \approx \frac{1}{A} \left( 1 - \frac{\lambda_0}{A} \right) \approx \frac{1}{A}.
\end{equation}
One can clearly see that  when $A$ is large, activation events cluster during peak phases, and this in  turn  significantly reduces inter-event intervals.

On the other hand,  for small oscillations  (where $A \ll \lambda_0$), one gets 
\begin{equation}
E[T_1] \approx t_w + \frac{1}{\lambda_{\text{eff}}} + \frac{A}{\lambda_0 \omega} \sin(\omega t_w).
\end{equation}
The effective activation rate over an oscillation period is given by 
\begin{equation}
\lambda_{\text{eff}} = \frac{1}{T} \int_0^T \lambda(t) dt = \lambda_0.
\end{equation}
The additional correction term reflects the periodic influences on the activation timing and  shows deviations from the homogeneous Poisson process.

The expected time for the $k$-th activation can be written as 
\begin{equation}
E[T_k] \approx t_w + \frac{k}{\lambda_{\text{eff}}} + \frac{A}{\lambda_0 \omega} \sin(\omega t_w)
\end{equation}
which  shows that  sinusoidal modulations alter the activation event spacing by  introducing periodic shifts in inter-event intervals.

To analyze event distributions across an oscillation cycle, let us now compute  the cycle-averaged waiting time 
\begin{equation}
E[T] \approx \frac{1}{T} \int_0^T \frac{1}{\lambda_0 + A \sin(\omega t)} dt.
\end{equation}
For large oscillations, an asymptotic expansion leads to
\begin{equation}
E[T] \approx \frac{\pi}{\omega A}.
\end{equation}
Clearly,  the waiting times  decrease  as $A$ increases. The above  approximation arises in the limit \( A \gg \lambda_0 \), where the activation rate \( \lambda(t) = \lambda_0 + A \sin(\omega t) \) becomes sharply peaked. In this regime, most activation events occur near the maxima of the sinusoid. The mean waiting time to the first event is then dominated by the interval around the first peak. Using a saddle-point or leading-order asymptotic argument, the expected activation time scales inversely with both the amplitude and frequency. 

As before, the viral load is given by $ V_t = v_0 \sum_{i=1}^{N_t} e^{g (t - T_i)} $.    After some algebra, we get the expectation of $V_t$ as
$ E[V_t] = v_0 \int_{t_w}^{t} \lambda(s) e^{g (t - s)} ds $ 
which can be further calculated as 
 \begin{widetext}
\begin{eqnarray} E[V_t] &=& 0, \quad t < t_w, \\ \nonumber
 &=& v_0 \lambda_0 \frac{1}{g} \left( e^{g (t - t_w)} - 1 \right)  + v_0 A \frac{g \sin(\omega t) - \omega \cos(\omega t) - e^{-g (t - t_w)} \left( g \sin(\omega t_w) - \omega \cos(\omega t_w) \right)}{g^2 + \omega^2}, \quad t \geq t_w. \end{eqnarray}
\end{widetext}
The variance of $V_t$  on other hand has a form 
\begin{equation}
\text{Var}(V_t) = v_0^2 \int_{t_w}^{t} \lambda(s) e^{2g (t - s)} ds.
\end{equation}
The above integral can be simplified to 
\begin{widetext}
\begin{eqnarray}
\text{Var}(V_t) &=& 0, \quad t < t_w, \\ \nonumber
 &=& v_0^2 \lambda_0 \frac{1}{2g} \left( e^{2g (t - t_w)} - 1 \right)  + v_0^2 A \frac{2g \sin(\omega t) - \omega \cos(\omega t) - e^{-2g (t - t_w)} \left( 2g \sin(\omega t_w) - \omega \cos(\omega t_w) \right)}{4g^2 + \omega^2}, \quad t \geq t_w.
\end{eqnarray}
\end{widetext}

For a periodically varying activation rate, the corresponding survival probability can be written as 
\begin{widetext}
\begin{equation}
S(t - t_w) = \exp\left[ -\lambda_0 (t - t_w) + \frac{A}{\omega} \left( \cos(\omega t) - \cos(\omega t_w) \right) \right]
\end{equation}
\end{widetext}

In the small-oscillation regime ($A \ll \lambda_0$), sinusoidal modulation induces only minor perturbations, and the effective activation rate remains $\lambda_0$  and this preserves  the near-constant inter-event time characteristic of a homogeneous Poisson process. Conversely, for large oscillations ($A \gg \lambda_0$), the activation becomes burst-like, with strong clustering during peaks and quiescence during troughs. In this regime, the expected waiting time is well-approximated by $E[T] \approx \pi / (\omega A)$  which shows that activation timing is governed by oscillation amplitude and frequency. This transition from Poisson-like to bursty dynamics has broad implications for systems such as neural spike trains, viral latency, and biochemical signaling, where phase-dependent clustering modulates functional outcomes.

To aid readability, we provide in Table~\ref{tab:variables} a summary of all variables and parameters used throughout the paper, including both physiological and mathematical quantities.
\begin{table}[htb]
\scriptsize  % Smaller font for two-column fit
\renewcommand{\arraystretch}{1.05}
\setlength{\tabcolsep}{3.8pt}
\centering
\caption{Summary of variables and parameters used in the model.}
\begin{tabular}{lll}
\hline\hline
Symbol & Description & Units \\
\hline
$t$ & Time & days \\
$\lambda(t)$ & Time-dependent activation rate & day$^{-1}$ \\
$\lambda_0$ & Baseline activation rate & day$^{-1}$ \\
$\Lambda(t)$ & Cumulative hazard function & -- \\
$A(t)$ & ART concentration & -- \\
$B(t)$ & LPA concentration & -- \\
$S(t)$ & LRA concentration & -- \\
$\omega$ & Frequency of periodic fluctuations & rad/day \\
$A$ & Amplitude of sinusoidal modulation & day$^{-1}$ \\
$\eta(t)$ & Gaussian noise (zero mean) & day$^{-1}$ \\
$\sigma_\lambda^2$ & Variance of $\eta(t)$ & day$^{-2}$ \\
$k_{\text{drug}}, K_{\text{drug}}$ & ART decay rate & day$^{-1}$ \\
$k_S$, $k_B$ & LRA and LPA decay rates & day$^{-1}$ \\
$v_0$ & Virus production per activated cell & virions \\
$g$ & Viral growth rate & day$^{-1}$ \\
$T_w$ & ART washout or delay time & days \\
$T_1$, $T_k$ & Waiting time for 1st/$k$-th reactivation & days \\
$S(t)$ & Survival probability & -- \\
$V(t)$ & Viral load & virions/mL \\
$L(t)$ & Latently infected cells & cells \\
$I(t)$ & Infected cells & cells \\
$T(t)$ & Target CD4$^+$ T cells & cells/$\mu$L \\
$\delta_L$ & Latent cell death rate & day$^{-1}$ \\
$\delta_I$ & Infected cell death rate & day$^{-1}$ \\
$\xi$ & Rate of latent cell formation & cells/(virion$\cdot$day) \\
$\beta$ & Rate of infected cell formation & cells/(virion$\cdot$day) \\
$p$ & Virus production rate & virions/(cell$\cdot$day) \\
$c$ & Viral clearance rate & day$^{-1}$ \\
$\epsilon$ & ART suppression coefficient & -- \\
$\gamma$ & LRA activation rate coefficient & day$^{-1}$ \\
$s$ & Source rate of target cells & cells/$\mu$L/day \\
$d$ & Death rate of target cells & day$^{-1}$ \\
$k$ & Target cell infection rate & mL/(virion$\cdot$day) \\
$\alpha$ & Immune stimulation rate & day$^{-1}$ \\
$\mu$ & Effector cell decay rate & day$^{-1}$ \\
$A_0$, $S_0$ & Initial ART/LRA concentrations & -- \\
$N_t$ & Number of activation events by time $t$ & count \\
\hline\hline
\end{tabular}
\label{tab:variables}
\end{table}

\subsection{Randomly Fluctuating Activation Rate}

In biological systems, randomly fluctuating activation rates provide a biologically important  framework for processes such as neural spiking, immune activation, and viral latency, where  in this case molecular noise and regulatory signals drive dynamic variability. Latent viruses (e.g., HIV and herpes) reactivate probabilistically based on the immune state and cellular stress. Similarly, gene expression and protein synthesis are governed by transcriptional and translational noise. Modeling such systems with fluctuating rates captures the full spectrum, from periodic control to noise-driven dynamics.

To account for stochastic variations inherent in biological and physical systems,  in  this section we consider  a Poisson process with a fluctuating activation rate 
\begin{equation}
\lambda(t) = \lambda_0 + \eta(t),
\end{equation}
We define \(\lambda(t)\) as the time-dependent activation rate of latent cells, modulated by drug concentrations or biological rhythms. The symbol \(\lambda_0\) refers to the baseline constant activation rate in the absence of external modulation. When the rate is constant over time, we simply write \(\lambda = \lambda_0\). $\eta(t)$ is a Gaussian noise process with zero mean and variance
\begin{equation}
E[\eta(t)] = 0, \quad \text{Var}[\eta(t)] = \sigma^2_{\lambda},
\end{equation}
Using the above equation, we study  the relation  of the inherent randomness in viral activation.

Since the expectation of the fluctuating rate is $E[\lambda(t)] = \lambda_0$, we investigate the impact of stochastic fluctuations on the expected waiting time between successive activations. The expected waiting time is given by
\begin{equation}
E[T_{n+1} - T_n] = E \left[ \frac{1}{\lambda(t)} \right],
\end{equation}
Taylor expanding  around $\lambda_0$  leads to
\begin{equation}
\frac{1}{\lambda(t)} = \frac{1}{\lambda_0 + \eta(t)} \approx \frac{1}{\lambda_0} \left( 1 - \frac{\eta(t)}{\lambda_0} + \frac{\eta^2(t)}{\lambda_0^2} \right).
\end{equation}
Since  $E[\eta(t)] = 0$, after some algebra we get 
\begin{equation}
E \left[ \frac{1}{\lambda(t)} \right] \approx \frac{1}{\lambda_0} \left( 1 + \frac{E[\eta^2(t)]}{\lambda_0^2} \right).
\end{equation}
Substituting $E[\eta^2(t)] = \sigma^2_{\lambda}$, the expected waiting time simplifies to
\begin{equation}
E[T] \approx \frac{1}{\lambda_0} \left( 1 + \frac{\sigma^2_{\lambda}}{\lambda_0^2} \right).
\end{equation}
These results  indicate   that   the presence of noise increases the expected waiting time between activation events in comparison to  the deterministic case. The magnitude of this increase depends on the ratio $\sigma^2_{\lambda}/\lambda_0^2$  which shows that stronger fluctuations prolong activation intervals.

For large fluctuations, where $\sigma^2_{\lambda} \gg \lambda_0^2$, the waiting time is further approximated as
\begin{equation}
E[T] \approx \frac{\sigma^2_{\lambda}}{\lambda_0^3}.
\end{equation}
This regime is particularly relevant in biological systems, where activation events are dominated by external noise sources such as gene regulatory networks. In this case,  transcription factor binding is probabilistic. Experimental studies suggest that noise-driven gene activation events can occur at rates as low as $10^{-4}$ per second,  which shows that  significant variability leads to cell-to-cell heterogeneity. Thus, the derived expression provides a quantitative framework for understanding how fluctuations shape the activation timing across diverse stochastic processes.

We  further justify this result  using the probability distribution of the cumulative rate function. The probability density   can be approximated as
\begin{equation}
P(\Lambda) \approx \frac{1}{\sqrt{2\pi \sigma^2_{\Lambda}}} \exp \left( -\frac{(\Lambda - \lambda_0 (t - t_w))^2}{2 \sigma^2_{\Lambda}} \right).
\end{equation}
after some algebra, we  calculate  the firing time  as
\begin{equation}
E[T_{n+1} - T_n] = \int_{0}^{\infty} t P(\Lambda) dt,
\end{equation}
Via  integration by parts, we get
\begin{equation}
E[T_{n+1} - T_n] = \int_{0}^{\infty} t \frac{1}{\sqrt{2\pi \sigma^2_{\Lambda}}} \exp \left( -\frac{(t - \lambda_0 (t - t_w))^2}{2 \sigma^2_{\Lambda}} \right) dt,
\end{equation}
By changing variables and applying Gaussian integral approximations, one gets
\begin{equation}
E[T_{n+1} - T_n] \approx t_w + \frac{1}{\lambda_0} \left( 1 + \frac{\sigma^2_{\lambda}}{\lambda_0^2} \right),
\end{equation}
For large fluctuations, this reduces to
\begin{equation}
E[T] \approx \frac{\sigma^2_{\lambda}}{\lambda_0^3}.
\end{equation}
Based on fluctuation regimes, these results can be categorized based on fluctuation regimes. For small fluctuations ( $\sigma^2_{\lambda} \ll \lambda_0^2$), the waiting time remains nearly constant. We can approximate   $E[T] \approx 1/\lambda_0$. In the case of moderate fluctuations, a quadratic correction emerges.  This modifies   the waiting time to  $E[T] \approx (1/\lambda_0) (1 + \sigma^2_{\lambda}/\lambda_0^2)$. For large fluctuations ( $\sigma^2_{\lambda} \gg \lambda_0^2$), noise effects dominate, leading to a waiting time approximation of $E[T] \approx \sigma^2_{\lambda}/\lambda_0^3$.

For exponentially growing viral load $
V_t = v_0 \sum_{i=1}^{N_t} e^{g (t - T_i)}
$,
We  calculate the  expectation of $ V_t $ as 
\begin{eqnarray} E[V_t] &=& 0, \quad t < t_w, \\ \nonumber 
 &=& v_0 \lambda_0  \frac{1}{g} \left( e^{g(t - t_w)} - 1 \right)
\left( 1 + \frac{\sigma^2_{\lambda} g}{(\lambda_0 - g)^3} \right), \quad t \geq t_w.
 \end{eqnarray}
Similarly, the variance of $ V_t $ is given by
\begin{eqnarray} 
\text{Var}[V_t] &=& 0, \quad t < t_w, \\ \nonumber
&=& v_0^2 \lambda_0 \frac{e^{2g (t - T_w)} - 1}{2g} \left( 1 + \frac{2 \sigma^2_{\lambda} g}{(\lambda_0 - g)^3} \right), \quad t \geq t_w.
\end{eqnarray}
These results indicate  the role of stochastic fluctuations  on the  viral load and its variance. The presence of noise introduces additional corrections and it  alters  the system's behavior as a function of the fluctuation intensity. 
For small fluctuations, where $ \sigma^2_{\lambda} \ll \lambda_0^2 $, the viral load follows a nearly deterministic trajectory, with deviations governed by the baseline activation rate $ \lambda_0 $. However, for large fluctuations ($ \sigma^2_{\lambda} \gg \lambda_0^2 $), noise dominates and this leads  to an increase   in both the expected viral load and its variance.

The survival probability is given as
\begin{equation}
S(t - t_w) \approx \exp \left( -\lambda_0 (t - t_w) + \frac{\sigma^2_{\lambda}}{2} (t - t_w)^2 \right).
\end{equation}
These results indicate  that  as the fluctuation intensity increases, systems transition from near-deterministic behavior to regimes dominated by stochastic dynamics.

\subsection{The role  of Pharmacokinetic Decay on Latent HIV Activation: A Mathematical Analysis}

Antiretroviral therapy (ART) effectively suppresses HIV replication through combination regimens targeting multiple stages of the viral life cycle. The pharmacokinetics of these agents involve complex metabolism and transporter interactions that can influence treatment efficacy and reactivation risk \cite{yu2023}.

The activation rate of latent cells following ART interruption is dictated by the pharmacokinetics of drug clearance. To model this effect, we define the time-dependent activation rate as
\begin{equation}
\lambda(t) = \lambda_0 (1 -   e^{-k_{\text{drug}} t}).
\end{equation}
Here, $\lambda_0$ and $k_{\text{drug}}$   denote the baseline activation rate in the absence of ART and  the drug elimination rate, respectively. At $t = 0$, the activation rate is fully suppressed ($\lambda(0) = 0$). As the drug concentration declines exponentially over time, the activation rate gradually increases.  As time progresses,  it approaches  its  maximum value, $\lambda_0$ as $t \to \infty$.

The rate at which reactivation occurs critically depends on parameter $k_{\text{drug}}$. When  $k_{\text{drug}}$ is small, drug clearance  becomes slow.  This leads to prolonged suppression of viral reactivation.  On the other hand, for large  $k_{\text{drug}}$, drug elimination becomes fast.  This, in turn,  accelerates   the resurgence of viral replication.  These findings also reflect  that   both  $\lambda_0$ and $k_{\text{drug}}$  play a crucial role in determining the timing and extent of viral resurgence since they  provide a quantitative framework for studying post-treatment control and optimizing therapeutic strategies.
Next to  model the decay of antiretroviral therapy (ART) concentration over time, let us   assume an exponentially  decreasing function 
\begin{equation}
A(t) =   e^{-k_{\text{drug}} t}
\end{equation}
where  $k_{\text{drug}}$  denotes  drug elimination rate, respectively. We show that as ART concentration declines, viral replication and latent cell activation gradually resume.

To analyze the impact of pharmacokinetic decay on HIV latent cell activation, we model the cumulative hazard function as $
\Lambda(t) = \int_0^t \lambda_0 \left(1 - e^{-k_{\text{drug}} \tau} \right) d\tau.
$
After some algebra, we get 
$
\Lambda(t) = \lambda_0 \left[t - \frac{1 - e^{-k_{\text{drug}} t}}{k_{\text{drug}}} \right].
$

The expected activation time is given by
$
E[T_1] = \int_0^\infty e^{-\Lambda(t)} dt,
$
For \textbf{small} $k_{\text{drug}}$, expanding the exponential term in $\Lambda(t)$ and solving perturbatively,
 \begin{equation}
        E[T_1] \approx \sqrt{\frac{\pi}{2 \lambda_0 k_{\text{drug}}}}.
    \end{equation}
This clearly depicts   delayed activation due to prolonged drug presence.
For \textbf{large} $k_{\text{drug}}$, rapid drug clearance leads to
\begin{equation}
E[T_1] \approx \frac{1}{\lambda_0}.
\end{equation}
This also indicates  that  for sufficiently fast drug decay, the activation time converges to its Poissonian limit.

For moderate $k_{\text{drug}}$, retaining dominant correction terms, one gets 
\begin{equation}
        E[T_1] \approx \frac{1}{\lambda_0} + \sqrt{\frac{\pi}{2}} \frac{1}{\sqrt{\lambda_0 k_{\text{drug}}}},
    \end{equation}
	This result highlights the diminishing influence of the drug clearance rate beyond a certain threshold.
This analysis shows that while drug presence initially delays activation, its long-term impact diminishes as clearance accelerates. The fundamental activation timescale remains governed by $\lambda_0$. This also implies  that pharmacokinetic decay must be considered in therapeutic strategies but does not fundamentally alter the stochastic nature of latency reversal.

After some algebra, the expectation of $ V_t $ is given by
\begin{widetext}
\begin{eqnarray}
E[V_t] &=& 0, \quad t < t_w, \\ \nonumber
 &=& v_0 \lambda_0 \left[t - \frac{1 - e^{-k_{\text{drug}} t}}{k_{\text{drug}}} \right] \frac{1}{gt} \left( e^{g(t - t_w)} - 1 \right)  
\left( 1 - \frac{  k_{\text{drug}}}{(\lambda_0 - g)(\lambda_0 - g + k_{\text{drug}})} e^{-k_{\text{drug}} t} \right), \quad t \geq t_w.
\end{eqnarray}
\end{widetext}
The contour plot (Fig, 1) of the expected viral load $E[V_t]$ as a function of the ART decay rate $k_{\mathrm{drug}}$ and time $t$ demonstrates that slower pharmacokinetic clearance markedly suppresses rebound. When $k_{\mathrm{drug}}$ is small, ART levels gradually decay, maintaining suppression, and delaying viral expansion. This effect arises from the sublinear growth of the activation term and the exponential correction due to the presence of the residual drug. Using the parameters $v_0 = 10^5$, $\lambda_0 = 0.2$,  $g = 0.3$, and $t_w = 5$, the model predicts a low-rebound regime consistent with effective post-treatment control. These findings stress  the importance of sustained ART in widening the shock-and-kill therapeutic window.

\begin{figure}[ht]
\centering
\includegraphics[width=0.4\textwidth]{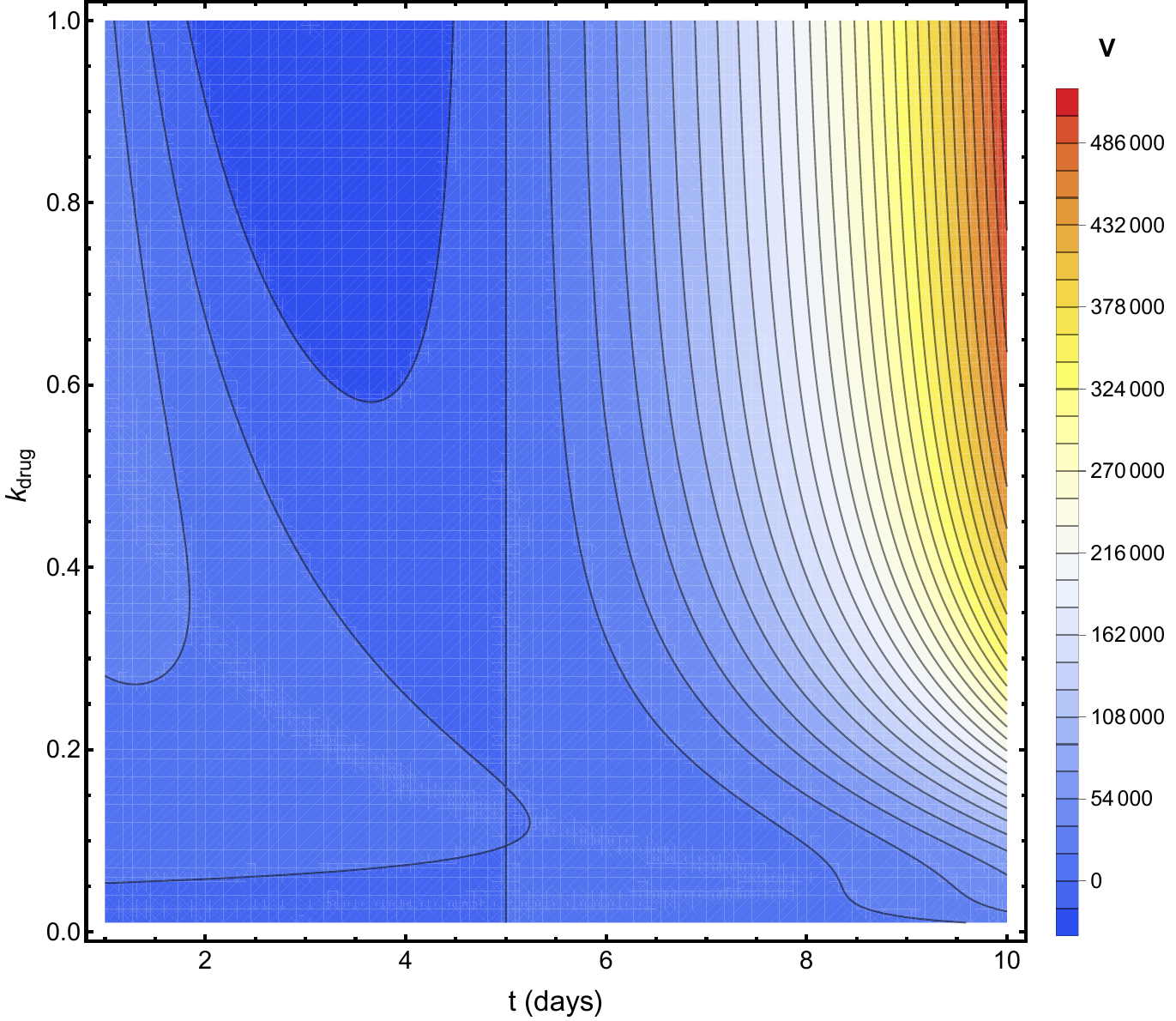}
\caption{
Contour plot of the expected viral load $E[V_t]$ as a function of the drug elimination rate $k_{\mathrm{drug}}$ and time $t$ based on analytical expression modeling of pharmacokinetic ART washout. As shown in the figure, when $k_{\mathrm{drug}}$ is low, ART decays slowly, maintaining the suppression of viral replication. This results in lower viral loads throughout the observed time window.  In the figure, we fix $v_0 = 10^5$, $\lambda_0 = 0.2$, $g = 0.3$, and $t_w = 5$. 
}
\label{fig:EV_Contour_ART}
\end{figure}

The survival probability after some algebra is given by 
\begin{equation}
S(t - t_w) = \exp \left[ -\lambda_0 \left( (t - t_w) - \frac{ }{k_{\text{drug}}} (1 - e^{-k_{\text{drug}} (t - t_w)}) \right) \right].
\end{equation}
In Fig.~2, the survival probability $S(t)$ is plotted  as a function of time $t$   for different activation rate models with waiting time $T_w = 2$. Before the   waiting time $T_w$, $S(t) = 1$  because  activation has not started. After the activation time  $T_w$, $S(t)$ decreases based on the activation model: constant rate, drug washout, sinusoidal activation, and stochastic fluctuations. The stochastic model introduces small perturbations around the baseline rate. The vertical dotted line indicates $T_w$. The other parameters  are fixed as  $\lambda_0 = 0.5$, $k_{\text{drug}} = 0.7$, $A = 0.2$, $\omega = \frac{2\pi}{5}$, and $\eta(t) \sim \mathcal{N}(0, 0.1)$. In 
Fig.~3, we plot the survival probability $S(t)$  as a function of time $t$ for  sinusoidal activation with varying amplitude $A$. A larger $A$ increases oscillations, affecting the survival probability.  In the figure, the other parameters are fixed as  $\lambda_0 = 0.5$, $\omega = \frac{2\pi}{5}$, and $A \in \{0.1, 0.3, 0.5\}$.
\begin{figure}[h]
    \centering
    \includegraphics[width=0.45\textwidth]{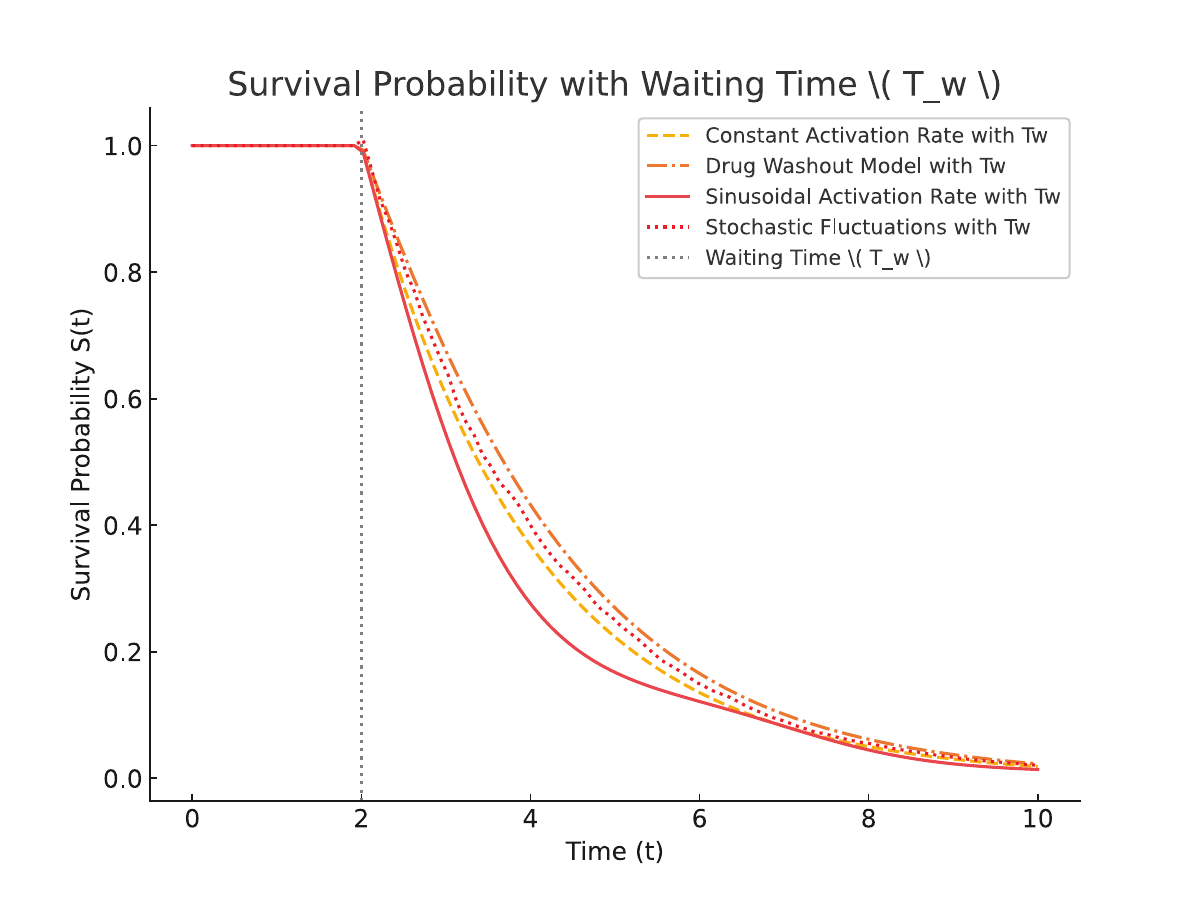}
    \caption{The survival probability $S(t)$ as a function of time under different activation rate models, incorporating a waiting time $T_w = 2$. Before $T_w$, the survival probability remains unity because activation has not commenced. After $T_w$, the probability decreases according to different models: constant activation rate, drug washout model, sinusoidal activation model, and stochastic fluctuation model. The stochastic model includes small perturbations around the baseline activation rate. The vertical dotted line indicates the waiting time $T_w$. The parameters used are $\lambda_0 = 0.5$ (baseline activation rate),  $k_{\text{drug}} = 0.7$ (drug decay rate), $A = 0.2$ (sinusoidal amplitude), $\omega = \frac{2\pi}{5}$ (sinusoidal frequency, corresponding to a period of 5), and stochastic fluctuations modeled as $\eta(t) \sim \mathcal{N}(0, 0.1)$ with mean zero and standard deviation of 0.1.}
    \label{fig:survival_tw}
\end{figure}

\begin{figure}[h]
    \centering
    \includegraphics[width=0.45\textwidth]{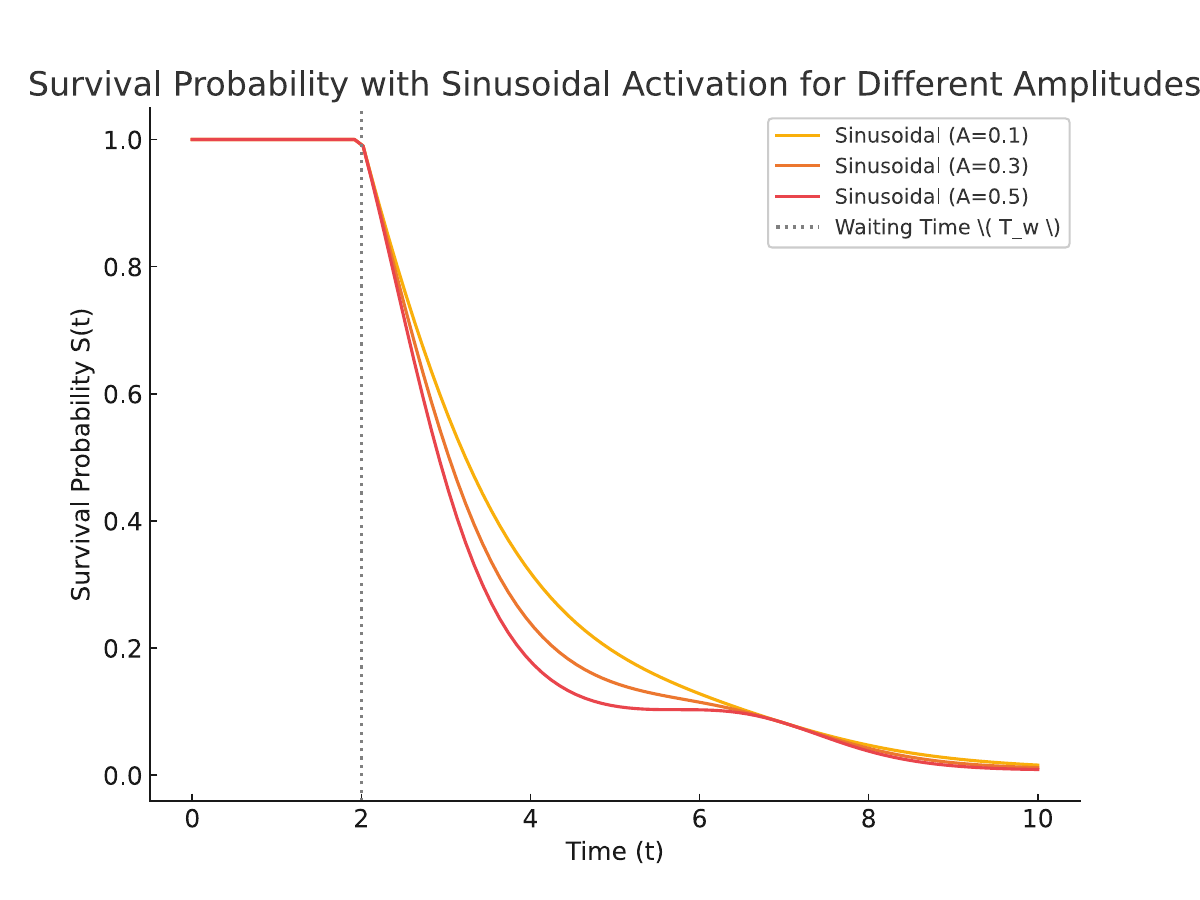}
    \caption{The survival probability $S(t)$  as  a function of time  $t$ for a sinusoidal activation rate model for different  $A$.  Here, we  fix  $T_w = 2$. The sinusoidal activation rate introduces periodic fluctuations in activation probability, which  affects  survival dynamics.  The vertical dotted line represents the waiting time $T_w$. We fixed other parameters as $\lambda_0 = 0.5$ (baseline activation rate), $\omega = \frac{2\pi}{5}$ (sinusoidal frequency, corresponding to a period of 5), and three different amplitudes $A = [0.1, 0.3, 0.5]$.}
    \label{fig:survival_sinusoidal}
\end{figure}

\subsection{The Role of Pharmacokinetic Decay with ART and LRA on Latent HIV Activation in Shock-and-Kill Strategies}

The ``Shock-and-Kill'' strategy aims to eliminate latent HIV by using latency-reversing agents (LRAs) to activate dormant viruses, followed by clearance via ART or the immune system. In one regime, LRAs and ART are applied together, which can reduce viral rebound by blocking reinfection but may also limit immune clearance due to weak antigen presentation and possible interference with LRA efficacy. In contrast, applying LRAs during a period of ART interruption can enhance immune recognition but carries the risk of uncontrolled replication, immune escape, and reservoir reseeding if clearance is delayed. To capture the interplay between reactivation and pharmacokinetics, we model the time-dependent activation rate by incorporating decaying concentrations of ART and LRA. Specifically, we define
\begin{equation}
\lambda(t) = \lambda_0 \left(1 - A(t) \right) + \gamma S(t),
\end{equation}
where \(A(t) =   e^{-K_{drug} t}\) and \(S(t) = S_0 e^{-k_S t}\) represent the pharmacokinetic decay of ART and LRA with respective rates \(K_{drug}\) and \(k_S\). Here, \(\lambda_0\) denotes the baseline reactivation rate in the absence of ART, and \(\gamma\) quantifies the enhancement of reactivation due to LRA stimulation. This formulation allows us to analyze the timing and likelihood of latent cell reactivation and to identify the pharmacodynamic conditions under which shock-and-kill protocols are most effective.

At $t = 0$, ART exerts maximal suppression while LRA is at peak efficacy, yielding a suppressed or possibly even elevated activation rate depending on the relative magnitude  $S_0$. As time progresses and both drugs decay, $\lambda(t)$ evolves dynamically, eventually approaching the baseline rate $\lambda_0$ as $A(t), S(t) \to 0$.

To quantify the timing of latent cell reactivation, we evaluate the cumulative hazard:
\begin{equation}
\Lambda(t) = \int_0^t \left[ \lambda_0 (1 -   e^{-K_{drug} \tau}) + \gamma S_0 e^{-k_S \tau} \right] d\tau.
\end{equation}
Carrying out the integration yields:
\begin{equation}
\Lambda(t) = \lambda_0 \left[t - \frac{ 1}{K_{drug}} (1 - e^{-K_{drug} t}) \right] + \frac{\gamma S_0}{k_S} (1 - e^{-k_S t}).
\end{equation}

The expected time of first activation is given by
\begin{equation}
E[T_1] = \int_0^\infty e^{-\Lambda(t)} dt.
\end{equation}

\textbf{Small decay rates:} For small $K_{drug}$ and $k_S$, prolonged drug presence suppresses activation. Using perturbation analysis, the expected time becomes:
\begin{equation}
E[T_1] \approx \sqrt{ \frac{\pi}{2 (\lambda_0   K_{drug} + \gamma S_0 k_S)} },
\end{equation}
which reflects a delay in reactivation due to slow clearance of both ART and LRA.

\textbf{Large decay rates:} For rapidly cleared drugs ($K_{drug}, k_S \gg \lambda_0$), we have
\begin{equation}
E[T_1] \approx \frac{1}{\lambda_0},
\end{equation}
which is the Poissonian limit, as ART and LRA rapidly wash out and $\lambda(t) \to \lambda_0$.

\textbf{Intermediate regime:} Retaining leading corrections gives
\begin{equation}
E[T_1] \approx \frac{1}{\lambda_0} + \sqrt{ \frac{\pi}{2} } \frac{1}{ \sqrt{ \lambda_0   K_{drug} + \gamma S_0 k_S } },
\end{equation}
highlighting that the reactivation delay is jointly shaped by ART and LRA pharmacokinetics.

To quantify viral output, we compute the expected viral load $E[V_t]$ for $t \geq t_w$:
\begin{widetext}
\begin{eqnarray}
E[V_t] &=& 0, \quad t < t_w, \\ \nonumber
 &=& v_0 \int_{t_w}^t \lambda(s) \left( e^{g (t - s)} - 1 \right) ds \\ \nonumber
&=& v_0 \left[ \lambda_0 \left( t - t_w - \frac{1 }{K_{drug}} (1 - e^{-K_{drug} (t - t_w)}) \right) + \frac{\gamma S_0}{k_S} (1 - e^{-k_S (t - t_w)}) \right] \cdot \frac{1}{g t} \left(e^{g(t - t_w)} - 1 \right).
\end{eqnarray}
\end{widetext}

\begin{figure}[ht]
\centering
\includegraphics[width=0.4\textwidth]{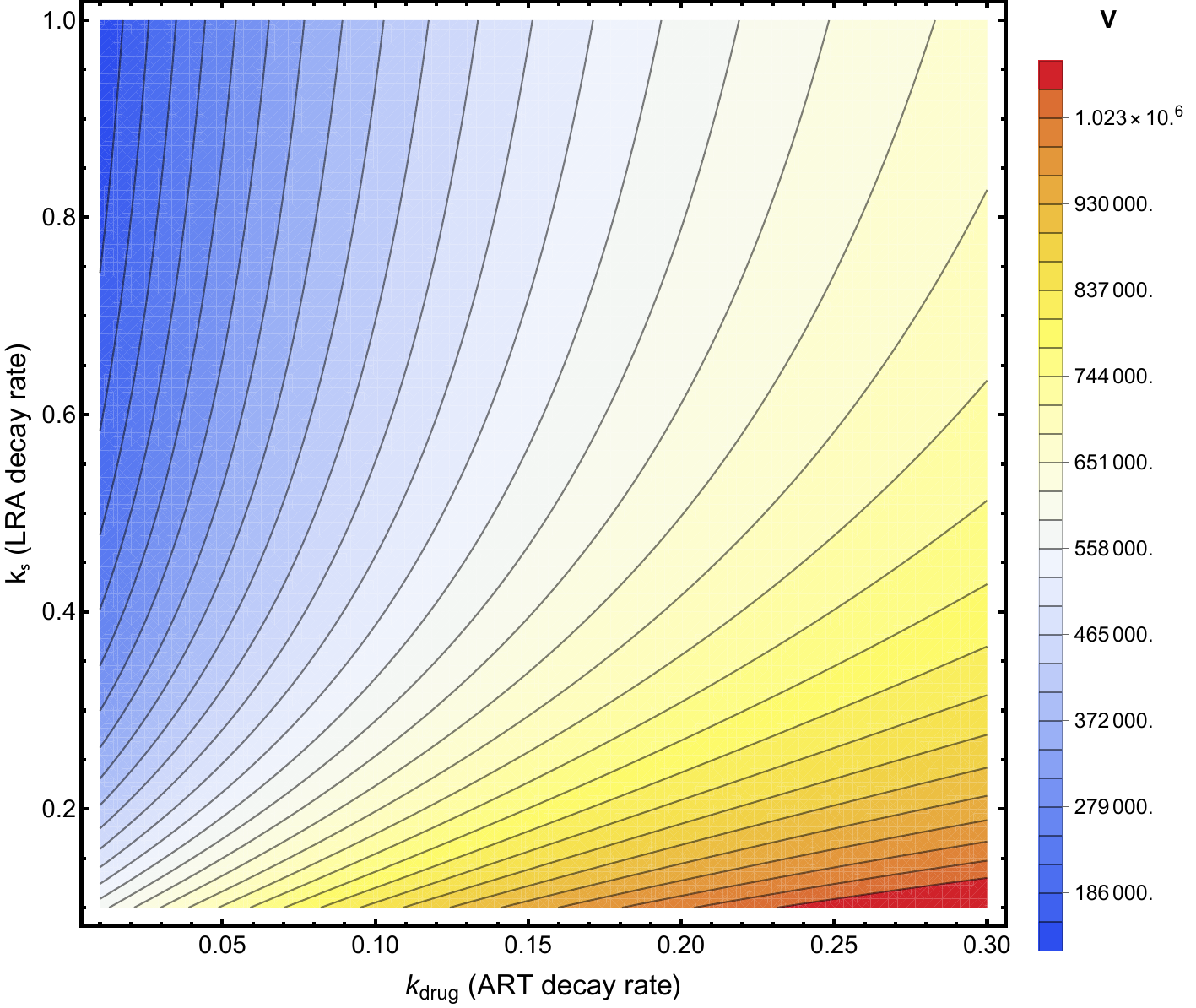}
\caption{
Contour plot of the expected viral load $E[V_t]$ at $t = 15$ days as a function of the ART decay rate $K_{drug}$ and LRA decay rate $k_S$. The plot is generated using a pharmacokinetically informed expression that incorporates time-varying activation from ART suppression and LRA stimulation. Parameter values used are $v_0 = 10^5$, $\lambda_0 = 0.2$, $S_0 = 1.0$, $\gamma = 0.2$, $g = 0.3$, and $t_w = 5$. The plot illustrates that slower ART decay (small $K_{drug}$) extends suppression, while faster LRA decay (large $k_S$) focuses reactivation into a narrow window, together enabling more effective post-treatment control through shock-and-kill.
}
\label{fig:EV_Contour}
\end{figure}
Figure 4  exhibits a distinct regime in which the expected viral load $E[V_t]$ is minimized as long as  the ART decay rate $K_{drug}$ is small and the LRA decay rate $k_S$ is large. This regime corresponds to the slow pharmacokinetic washout of ART that  sustains viral suppression, combined with a rapidly decaying LRA that delivers a sharp and transient activation signal. Such a configuration optimally aligns with the mechanistic requirements of the ``shock-and-kill'' strategy. Particurlaly, delayed ART clearance prevents premature reactivation at the  same time the transient LRA burst ensures timely latency reversal before immune exhaustion or viral rebound. In this parameter regime, viral reactivation is clustered into a controlled window, where immune or ART-mediated clearance is still active.  This, in turn, minimizes  the risk of systemic rebound and reservoir reseeding.  Our  findings provide quantitative support for the design of post-treatment control strategies that synchronize latency-reversing stimuli with pharmacodynamic decay and stress  the critical interplay between drug kinetics and reactivation timing in the efficacy of HIV cure interventions.

Finally, the survival probability—that is, the probability of no reactivation up to time $t \geq t_w$—is given by
\begin{widetext}
\begin{equation}
S(t - t_w) = \exp \left[ -\lambda_0 \left( (t - t_w) - \frac{1 }{K_{drug}} (1 - e^{-K_{drug} (t - t_w)}) \right) - \frac{\gamma S_0}{k_S} (1 - e^{-k_S (t - t_w)}) \right],
\end{equation}
\end{widetext}

This analysis demonstrates how the interplay between ART and LRA decay rates determines the timing and variability of latent cell reactivation.  It  also indicates that fast clearance of ART enhaene  activation and slow LRA decay can prolong immune exposure. Together, these dynamics provide a mathematical foundation for optimizing the timing and sequencing of ``shock-and-kill'' interventions.

The effectiveness of the shock-and-kill approach depends on a delicate balance between reactivating latent HIV and ensuring timely clearance before viral rebound occurs. We model the time-dependent activation rate as
$
\lambda(t) = \lambda_0 \left(1 -   e^{-K_{drug} t} \right) + \gamma S_0 e^{-k_S t},
$
where the ART and LRA decay rates \(K_{drug}\) and \(k_S\) control suppression and stimulation, respectively. The cumulative hazard \(\Lambda(t)\) quantifies the likelihood of reactivation, with likely activation when \(\Lambda(t) \gtrsim 1\). To ensure reactivation occurs during a therapeutically favorable window—after ART has decayed but LRA remains active—we require \(K_{drug} > k_S\) so that \(A(t) \ll 1\) and \(S(t) \gg 0\) overlap in time. The expected activation time 
$
\mathbb{E}[T_1] \approx \sqrt{\frac{\pi}{2(\lambda_0   K_{drug} + \gamma S_0 k_S)}}
$
must also fall within the window in which immune-mediated clearance remains viable. This defines a narrow but tunable opportunity for pharmacologically driven reactivation.

Following reactivation, the viral load grows as \(V_t = \sum_{i=1}^{N_t} v_0 e^{g(t - t_i)}\), where \(g\) is the replication rate. In order to prevent  reseeding of the latent reservoir, the expected viral burden must remain below a critical threshold 
$
\mathbb{E}[V_t] = v_0 \int_0^t \lambda(s) e^{g(t - s)} ds \ll V_{\text{reseed}}.
$
Alternatively, clearance dominates if the kill rate \(k_I > g\), ensuring the elimination of infected cells before significant viral expansion. Together, these conditions delineate the effective parameter space for shock-and-kill interventions and show how the success of this strategy is shaped by the tight interplay between drug kinetics, stochastic latency reversal, and post-activation immune- or ART-mediated control.

\begin{figure}[t]
\centering
\includegraphics[width=\linewidth]{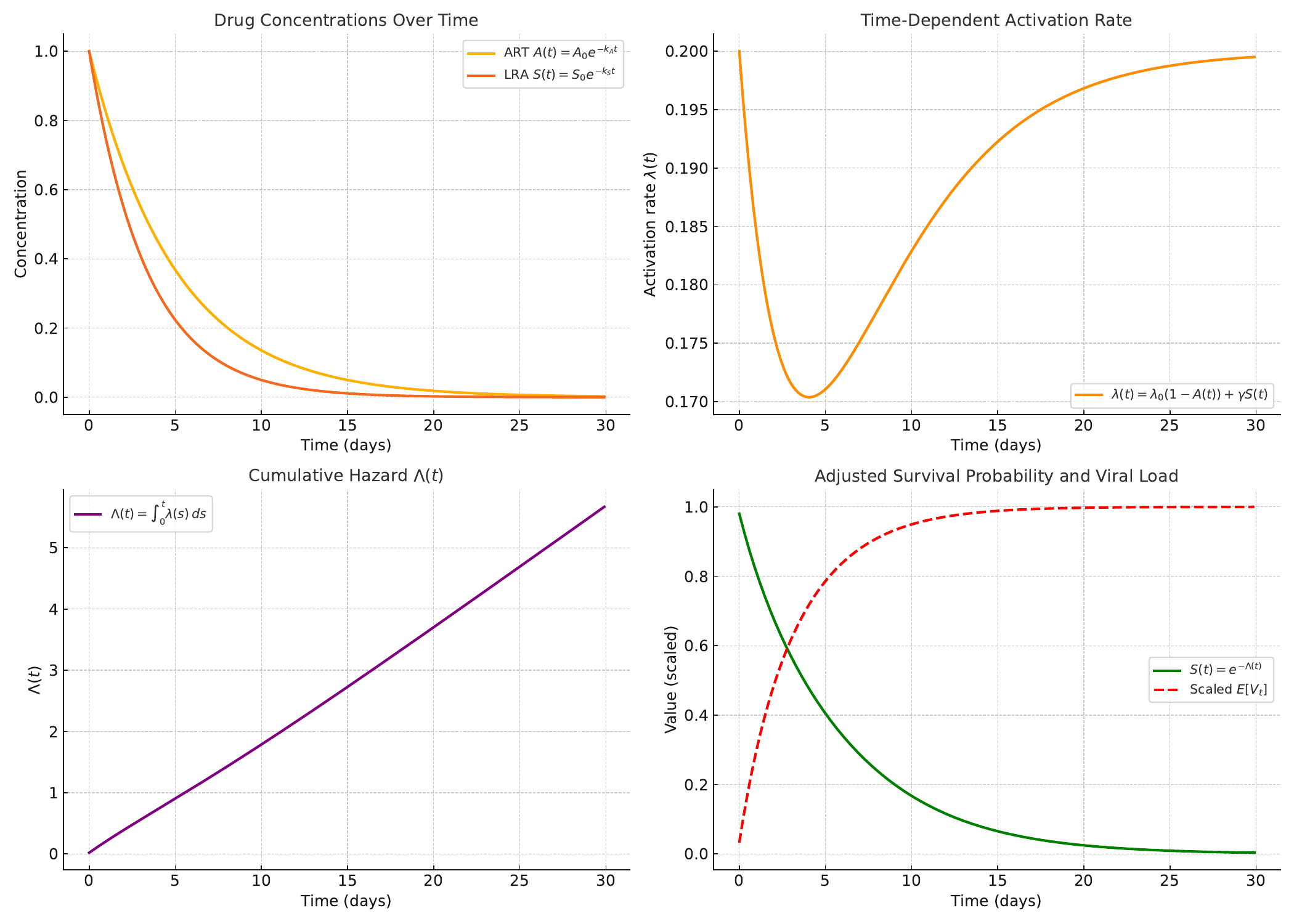}
\caption{
Dynamical features of a pharmacokinetically adjusted shock-and-kill protocol. (Top left) Drug concentrations decay exponentially: ART, $A(t) =   e^{-K_{drug} t}$, and LRA, $S(t) = S_0 e^{-k_S t}$, with $K_{drug} = 0.2$, $k_S = 0.3$. (Top right) reactivation rate $\lambda(t) = \lambda_0(1 - A(t)) + \gamma S(t)$ increases after ART decays and peaks during LRA activity. (Bottom left) cumulativetive hazard $\Lambda(t) = \int_0^t \lambda(s) ds$ determines the probability of latent cell reactivation. (Bottom right) Survival probability $S(t) = e^{-\Lambda(t)}$ declines gradually, while the scaled expected viral load $\mathbb{E}[V_t]/\max \mathbb{E}[V_t]$ reflects delayed rebound, which enables  immune clearance. Other parameters are fixed as  $\lambda_0 = 0.2$, $\gamma = 0.2$, $g = 0.3$, $v_0 = 1$.
}
\label{fig:shock-kill}
\end{figure}

Figure~\ref{fig:shock-kill} shows   how a carefully chosen pharmacokinetic regime enables an effective shock-and-kill strategy. By selecting a slower ART decay rate (\(K_{drug} = 0.2\)) and a faster LRA decay rate (\(k_S = 0.3\)), we create a therapeutic window in which ART suppression wanes while LRA stimulation remains active. This forces  the reactivation rate \(\lambda(t)\) to rise moderately and transiently,  which also avoids  early or excessive activation that could lead to uncontrolled rebound.

The cumulative hazard \(\Lambda(t)\) increases gradually, ensuring that reactivation is neither suppressed nor explosive. Correspondingly, the survival probability \(S(t)\) decays smoothly, and the expected viral load \(\mathbb{E}[V_t]\) grows with delay, giving the immune system or re-initiated ART sufficient time to eliminate reactivated cells. This parameter regime satisfies the criteria for effective shock (sufficient \(\Lambda(t)\)) and effective kill (delayed \(\mathbb{E}[V_t]\)), which  minimizes  the risk of viral rebound or reservoir reseeding.

\subsection{The Role of Pharmacokinetic Decay in Latency Silencing during Shock-and-Lock Strategies}

The ``Shock-and-Lock'' strategy aims to prevent HIV rebound by pharmacologically reinforcing viral latency. Unlike ``shock-and-kill'', which relies on latency reversal followed by clearance, this approach combines antiretroviral therapy (ART) with a latency-promoting agent (LPA) to suppress reactivation even after ART cessation. The combined pharmacokinetics of ART and the locking drug dynamically reduce the reactivation rate over time.

We define the time-dependent reactivation rate as
\begin{equation}
\lambda(t) = \lambda_0 (1 - A(t)) (1 -   B(t)),
\end{equation}
where \( A(t) = e^{-K_{drug} t} \) and \( B(t) = e^{-k_B t} \) describe the decaying concentrations of ART and LPA, respectively. 

The expected time to first reactivation, \( E[T_1] \), depends sensitively on the pharmacokinetic decay rates of ART (\( K_{\text{drug}} \)) and the latency-promoting agent (LPA, with decay rate \( k_B \)). Two asymptotic regimes can be analyzed analytically.
 When both drugs decay rapidly (\( K_{\text{drug}}, k_B \gg \lambda_0 \)), pharmacological suppression is short-lived and the reactivation rate quickly approaches its baseline value \( \lambda_0 \). In this limit, the expected time to reactivation reduces to the standard mean waiting time for a constant-rate Poisson process:
\begin{equation}
E[T_1] \approx \frac{1}{\lambda_0}.
\end{equation}
 In contrast, when both ART and LPA decay slowly (\( K_{\text{drug}}, k_B \ll \lambda_0 \)), pharmacological suppression persists over a prolonged timescale, and the reactivation rate increases gradually over time. In this limit, the rate behaves quadratically, \( \lambda(t) \sim t^2 \), leading to a cumulative hazard that scales as \( \Lambda(t) \sim t^3 \). The resulting survival probability takes the form of a stretched exponential, reflecting the strong temporal suppression of latency reversal. In this regime, the expected time to reactivation follows the scaling law:
\begin{equation}
E[T_1] \approx 2.67894 \cdot \left( 3 \lambda_0   K_{\text{drug}} k_B \right)^{-1/3}.
\end{equation}

This expression makes explicit the nonlinear dependence of reactivation timing on the pharmacokinetic parameters. Notably, increasing either \( k_B \) (faster decay of the locking agent) or \( K_{\text{drug}} \) (faster ART clearance) leads to earlier reactivation by narrowing the temporal window during which the latent reservoir is effectively suppressed. Conversely, slow elimination of both agents prolongs the suppressive state, thereby delaying reactivation. These results provide mechanistic insight into how pharmacological persistence extends viral quiescence and offer a quantitative foundation for optimizing drug half-lives in latency-control strategies..

To quantify the viral output conditional on reactivation, we compute the expected viral load for \( t \geq t_w \), where \( \tau = t - t_w \) denotes the time since the suppression window ended:
\begin{widetext}
\begin{eqnarray}
E[V_t] &=& \frac{v_0 \lambda_0}{g} (e^{g \tau} - 1 - g \tau) \nonumber \\
&& - v_0 \lambda_0 e^{-K_{drug} t} \left[ \frac{e^{(K_{drug}+g)\tau} - 1}{K_{drug} + g} - \frac{e^{K_{drug} \tau} - 1}{K_{drug}} \right] \nonumber \\
&& - v_0 \lambda_0   e^{-k_B t} \left[ \frac{e^{(k_B+g)\tau} - 1}{k_B + g} - \frac{e^{k_B \tau} - 1}{k_B} \right] \nonumber \\
&& + v_0 \lambda_0   e^{-(K_{drug} + k_B)t} \left[ \frac{e^{(K_{drug} + k_B + g)\tau} - 1}{K_{drug} + k_B + g} - \frac{e^{(K_{drug} + k_B)\tau} - 1}{K_{drug} + k_B} \right].
\end{eqnarray}
\end{widetext}

This expression characterizes the rebound magnitude shaped by viral growth and pharmacokinetically driven suppression. The time course of \( E[V_t] \) depends nonlinearly on \( K_{drug} \) and  \( k_B \),  capturing how drug decay kinetics and silencing efficacy jointly influence rebound dynamics.
\begin{figure}[h]
\centering
\includegraphics[width=0.4\textwidth]{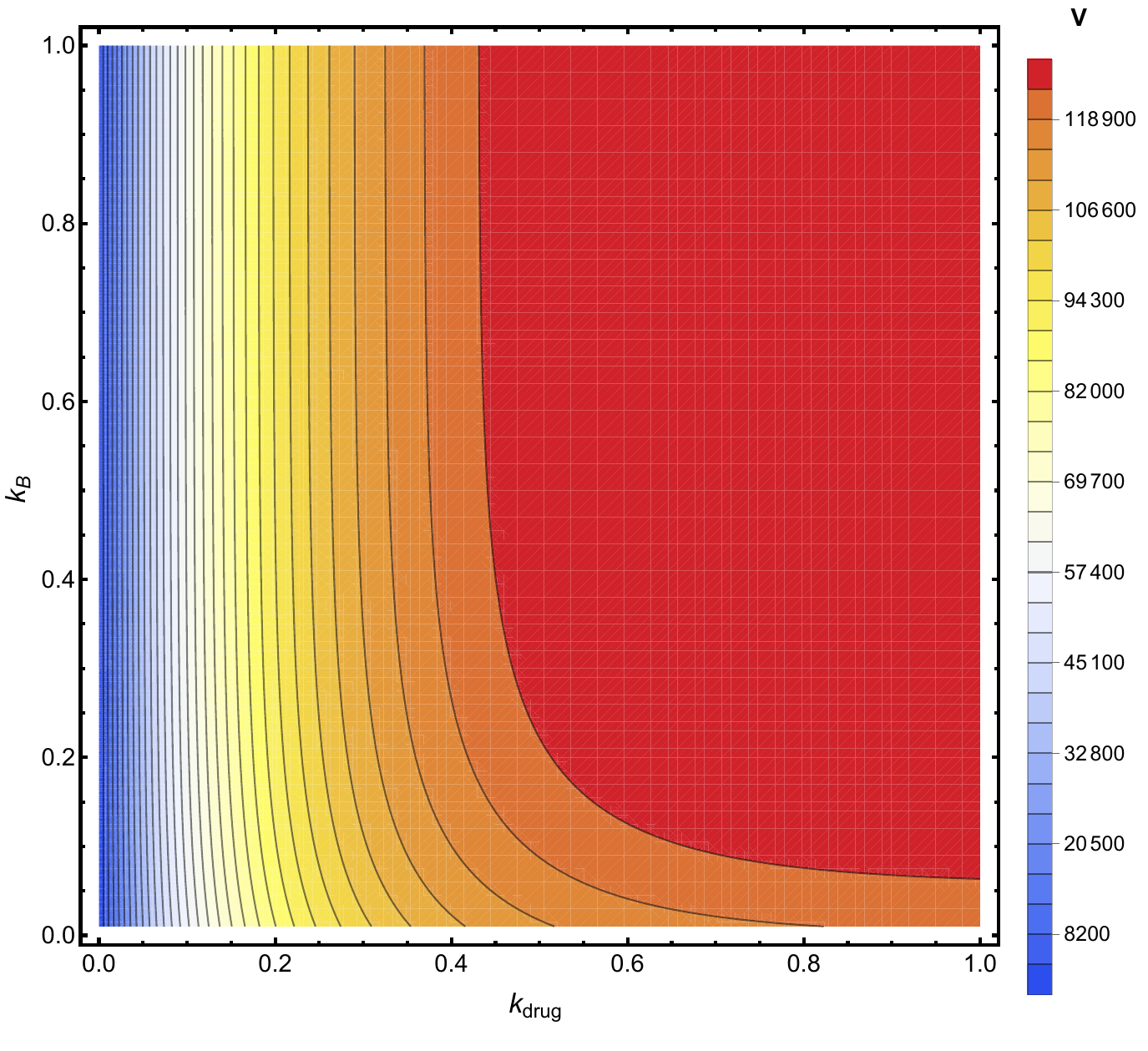}
\caption{ Contour plot of the expected viral load \( E[V_t] \) at fixed time \( t = 10 \) days, as a function of the ART decay rate \( k_{\text{drug}} \) (x-axis) and the LPA decay rate \( k_B \) (y-axis), assuming  \( t_w = 5 \). Darker regions indicate stronger suppression of viral rebound. The calculation uses \( \lambda_0 = 0.2 \), \( g = 0.3 \), and initial viral production rate \( v_0 = 10^5 \).}
\end{figure}
As shown in Fig.~6, the average viral load increases with both \( K_{\text{drug}} \) and \( k_B \). This behavior arises because faster decay of ART and the latency-promoting agent shortens the duration of pharmacological suppression. As drug levels fall more quickly, the reactivation rate rises sooner, allowing the virus to rebound earlier and grow over a longer time window. Consequently, the total viral output by a fixed time point is higher. These results emphasize the importance of sustained drug exposure for maintaining latency and minimizing rebound magnitude.

\section{Modeling HIV Reactivation with a Gamma-Distributed Waiting Time}

Traditional models of HIV latency assume exponentially distributed waiting times, implying Poissonian reactivation with constant hazard rates. However, biological evidence indicates significant heterogeneity across infected cells, suggesting that exponential assumptions may oversimplify latency dynamics. In this work, we derive analytical expressions for the expected activation time \( E[T] \) and cumulative reactivations \( E[N_t] \) under Gamma-distributed waiting times, extending classical Poisson models to a more flexible and biologically realistic framework. 

Let now assume that the waiting time until HIV reactivation follows a Gamma distribution with shape parameter $k$ and rate parameter $\theta$. The probability density function  can be written as 
\begin{equation}
    f(T) = \frac{\theta^k T^{k-1} e^{-\theta T}}{\Gamma(k)}, \quad T \geq 0.
\end{equation}
The cumulative distribution function is give by 
\begin{equation}
    F(T) = \frac{\gamma(k, \theta T)}{\Gamma(k)},
\end{equation}
where $\gamma(k, \theta T)$  denotes the lower incomplete Gamma function. The expected activation time  has a form
\begin{equation}
    E[T] = \frac{k}{\theta}.
\end{equation}
For $k = 1$, the Gamma distribution reduces to the exponential distribution which recovers  the Poisson process,
\begin{equation}
    \lim_{k \to 1} E[T] = \frac{1}{\theta}.
\end{equation}

In this section, we consider the   previous model system where the  drug concentration exponentially decays.   The expected activation time $E[T_1]$ depends on the pharmacokinetic decay rate $k_{\text{drug}}$. For small $k_{\text{drug}}$, one gets
$
    E[T_1] \approx \sqrt{\frac{\pi}{2 \lambda_0 k_{\text{drug}}}}.
$
For moderate $k_{\text{drug}}$, retaining dominant correction terms,
$
    E[T_1] \approx \frac{1}{\lambda_0} + \sqrt{\frac{\pi}{2}} \frac{1}{\sqrt{\lambda_0 k_{\text{drug}}}}.
$
For large $k_{\text{drug}}$ (where drug clearance is rapid), we get 
$
    E[T_1] \approx \frac{1}{\lambda_0}
$.

For $k > 1$, activation times are more spread out and this increases  the variability in latent cell reactivation. The expected number of reactivations up to time $t$ is given as 
\begin{equation}
    E[N_t] = \lambda_0 \int_0^t F(\tau) d\tau.
\end{equation}
After some algebra, one gets 
\begin{equation}
    E[N_t] = \lambda_0 \int_0^t \frac{\gamma(k, \theta (t - \tau))}{\Gamma(k)} d\tau.
\end{equation}
For large $t$, the lower incomplete Gamma function satisfies the asymptotic relation
\begin{equation}
    \gamma(k, \theta t) \approx \Gamma(k), \quad \text{for } t \gg k/\theta.
\end{equation}
Approximating for large $t$,
\begin{equation}
    E[N_t] \approx \lambda_0 \left[ t - \frac{A_0}{k_{\text{drug}}} (1 - e^{-k_{\text{drug}} t}) \right].
\end{equation}
As $k \to 1$, the Poisson model is recovered, confirming that the Gamma process generalizes the Poisson case.

In the next section, we consider a deterministic model and explore the interaction of latent cells, free virus, and infected cells. We introduce a pharmacokinetic model to study the dynamics of viral reactivation. The also study the impact of immune response that  plays a critical role in controlling viral reactivation. Furthermore, we examine the "Shock-and-Kill" strategy, a promising approach for eradicating latent HIV reservoirs by strategically coordinating latency reversal agents (LRAs) with antiretroviral therapy (ART).

\section{Deterministic Modeling of Viral Dynamics Following Drug Washout}

Now, let us study  the dynamics of the system  during the drug washout.   
Before drug washout, antiretroviral therapy (ART) effectively hinders  viral replication. This prevents new infections and latent cell activation. 
The latent cell population remains constant since ART prevents activation
\begin{equation}
\frac{dL}{dt} = 0.
\end{equation}
Similarly, the population of infected cells is stabilized as ART inhibits viral replication and the formation of new infections
\begin{equation}
\frac{dI}{dt} = 0.
\end{equation}
The target cell population follows natural regeneration and decay in the absence of significant viral interactions
\begin{equation}
\frac{dT}{dt} = s - dT.
\end{equation}
The viral load remains at a suppressed level since ART prevents active replication.
\begin{equation}
\frac{dV}{dt} = 0.
\end{equation}

Next, we study  the dynamics after drug washout ($t \geq t_w$). Suppression of viral replication leads to viral rebound. As a result, latent cells become activated, and more host cells become infected. To account for ART decay over time, we introduce a pharmacokinetic model for drug clearance as before  
$
A(t) = A_0 e^{-k_{\text{drug}} t},
$
where $A_0$, $k_{\text{drug}}$ and $\lambda$ denote the initial drug concentration, drug elimination rate, and activation rate of the latent cells, respectively. The activation rate depends on ART as 
$
\lambda(t) = \lambda_0 (1 - A_0 e^{-k_{\text{drug}} t})
$.
Thus, the latent cell population decreases as a result of activation. It is also replenished by new latently infected cells formed through viral interactions
\begin{equation}
\frac{dL}{dt} = -\lambda_0 L (1 - A_0 e^{-k_{\text{drug}} t}) + \xi V (1 - A_0 e^{-k_{\text{drug}} t}) - \delta_L L.
\end{equation}
The infected cell population increases as latent cells reactivate, and new target cells become infected. As a result 
\begin{equation}
\frac{dI}{dt} = \lambda_0 L (1 - A_0 e^{-k_{\text{drug}} t}) + k T V (1 - A_0 e^{-k_{\text{drug}} t}) - \delta_I I.
\end{equation}
Target cells continue to regenerate naturally but are depleted due to infection by the virus. Hence 
\begin{equation}
\frac{dT}{dt} = s - dT - k T V.
\end{equation}
The viral load increases as infected cells produce new virions at the same time  that the clearance rate occurs at a constant rate. Therefore, the dynamics of the virus is given by
\begin{equation}
\frac{dV}{dt} = p I (1 - A_0 e^{-k_{\text{drug}} t}) - c V.
\end{equation}
Here, $\xi$  denotes  the rate at which new latent cells are formed through viral infection, and $\delta_L$ and $\delta_I$  represent  the natural death rates of latently infected and productively infected cells, respectively. The parameter $k$ represents the infection rate of target cells by the virus, $s$ is the source rate of new target cells, and $d$ is their natural death rate. Parameter $p$ is the rate at which infected cells produce new virions and $c$ is the clearance rate of free virus particles. The variables $T$, $V$, $I$, and $L$ represent the populations of the target cells, free virus, infected cells, and latently infected cells, respectively.

Before drug washout, ART maintains viral suppression by  preventing latent cell activation and development of new infections.  As a result,   the system becomes stable. However, after ART cessation, the drug concentration decreases over time, which  progressively allows  viral replication to resume. This results in reactivation of latent cells,  elevated infection levels, and eventual viral resurgence. The presence  of ART decay via $A(t) = A_0 e^{-k_{\text{drug}} t}$ leads to 
 a more accurate representation of post-treatment dynamics. This  decay  indicates   a gradual transition from viral suppression to active infection. The presence  of time-dependent $\lambda(t)$ accounts for the varying activation rates of latent cells based on ART availability, improving the model's biological relevance.

Before drug washout, antiretroviral therapy (ART) effectively suppresses viral replication  because it prevents   new infections as well as  latent cell activation.  At equilibrium , we get  
\begin{eqnarray} 
L^* & = & L_0, \\
I^* & = & I_0, \\
T^* & = & \frac{s}{d}, \\
V^* & = &V_ 0.
\end{eqnarray} 
These equations emphasize the  stabilizing effect of ART. When latent cells remain inactive, infection rates  are  suppressed. Any interruption risks disrupting this balance, which in turn  leads  to viral rebound.

After ART is  stopped  , we write  relation  at equilibrium as 
\begin{eqnarray} 
L^* &=& \frac{\xi V^*}{\lambda_0+ \delta_L}, \\ 
I^* &=& \frac{\lambda_0 L^* + k T^* V^*}{\delta_I}, \\ 
T^* &=& \frac{s}{d + kV^*}, \\ 
V^* &=& \frac{p I^*}{c}.
\end{eqnarray}These equilibrium conditions reveal that ART cessation triggers viral rebound, latent cell activation, and immune destabilization—highlighting the fragility of post-treatment control.

The rate of viral rebound is governed by the decay of ART concentration, \( A_0 e^{-k_{\text{drug}} t} \) and this stresses   the importance of pharmacokinetics in treatment design. Rapid ART withdrawal may accelerate  rebound. However,  gradual  cycling can dampen reactivation by limiting the availability of the target cells. The timing of LRA administration relative to viral load dynamics is also critical, and maximal reservoir reduction occurs when \( V^* \) remains low and this  minimizes  the risk of reseeding. Moreover, the dependence of viral and infected cell equilibria on ART decay suggests structured interruptions in delaying viral  rebound at the same time  reducing drug burden. Overall, equilibrium analysis before and after drug washout reveals how ART stabilizes latency. This  also shows  how its removal drives reactivation via viral replication, target cell depletion, and immune disruption. Hence, introducing  time-dependent ART decay is vital  because  it provides a biologically realistic framework for optimizing HIV cure strategies.

\begin{figure}[h]
    \centering
    \includegraphics[width=0.5\textwidth]{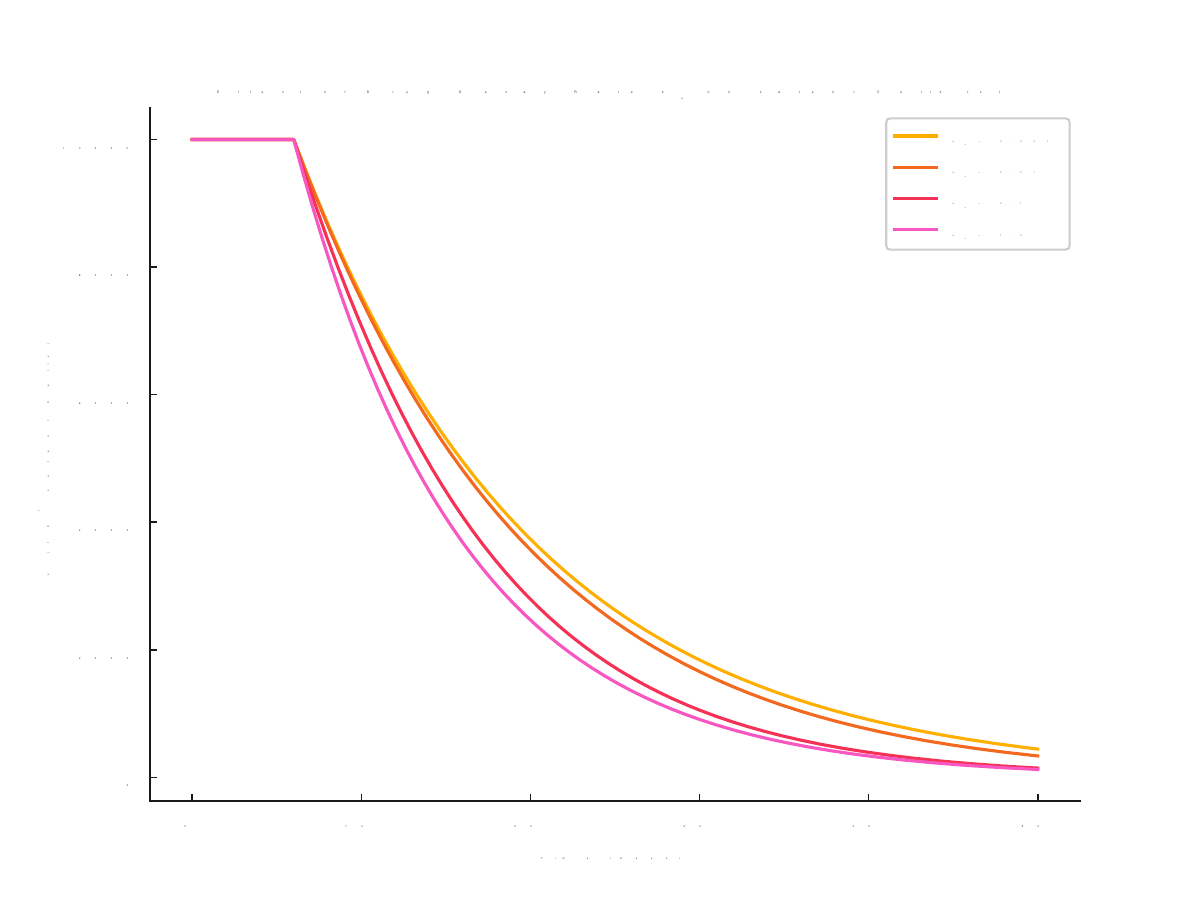}
    \caption{Time evolution of latent cells, $L(t)$, for different drug decay rates, $k_d = 0.001, 0.01, 0.1, 1.0$  for fixed parameters  $s = 100000$, $\lambda = 1 \times 10^{-6}$, $d = 0.1$, $h = 0.1$, 
    $e = 1 \times 10^{-10}$, $d_1 = 1.0$, $p_1 = 10^5$, $c_1 = 29$, $A_0 = 0.3$, and $T_0 = 10^5$.}
    \label{fig:latent_cells_kd}
\end{figure}

\begin{figure}[h]
    \centering
    \includegraphics[width=0.5\textwidth]{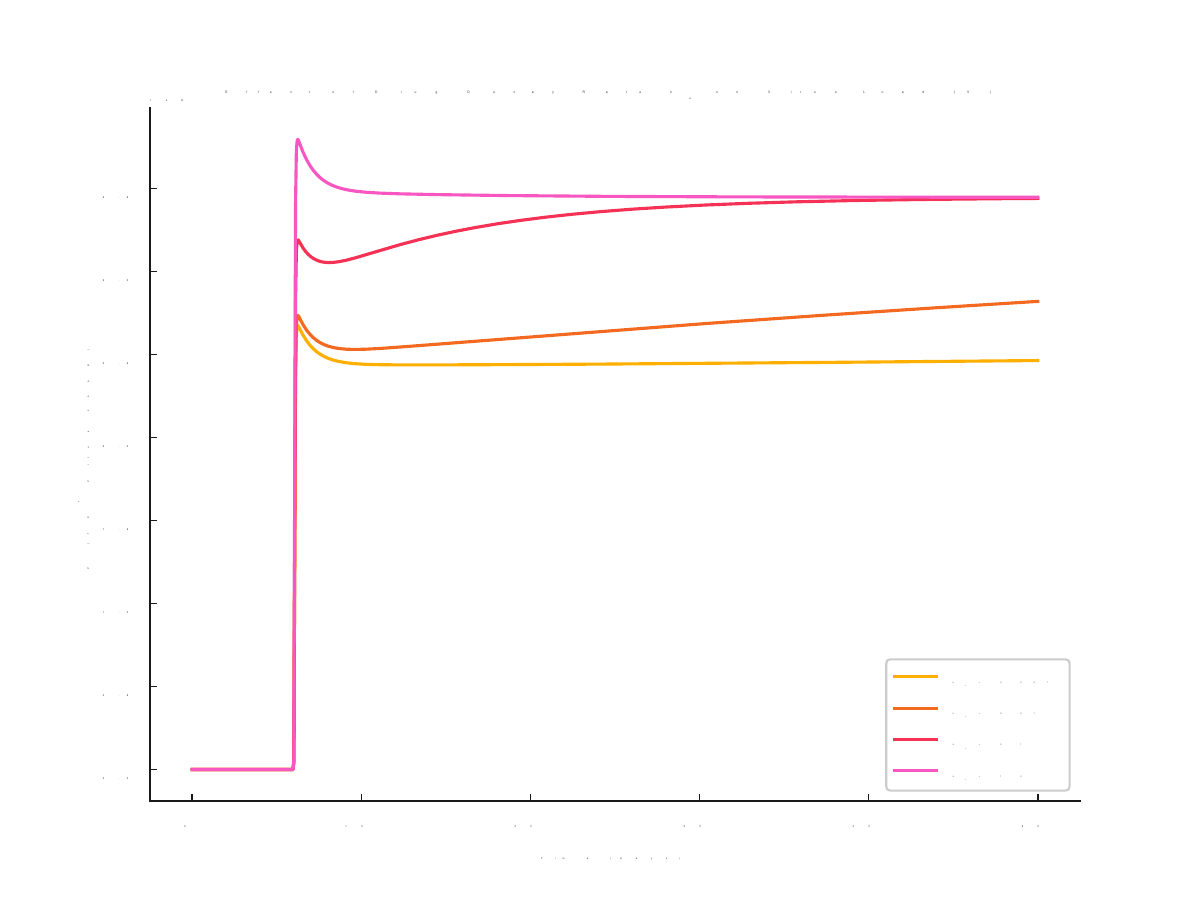}
    \caption{The dependence of  the virus load, $V(t)$ as function of time  for different  drug decay rates, $k_d = 0.001, 0.01, 0.1, 1.0$. 
    The viral load is significantly affected by the rate at which the drug decays, leading to different viral clearance or persistence dynamics. 
    The parameters used in the model are: $s = 100000$, $\lambda = 1 \times 10^{-6}$, $d = 0.1$, $h = 0.1$, 
    $e = 1 \times 10^{-10}$, $d_1 = 1.0$, $p_1 = 10^5$, $c_1 = 29$, $A_0 = 0.3$, and $T_0 = 10^5$.}
    \label{fig:virus_kd}
\end{figure}
    Let us now   explore this model system.  In  Fig. 7, we plot 		
the  time evolution of latent cells, $L(t)$ as  a function of time  for different drug decay rates, $k_d = 0.001, 0.01, 0.1, 1.0$. These results illustrate that as the drug washes out more slowly, the relaxation time for latent cells is extended by  delaying activation. Fig.~8 depicts $V(t)$  as a function of time $t$ for  the same drug decay rate, as shown in Fig. 7. The rate at which the drug decays significantly affects viral clearance and persistence. Prolonged drug washout results in a substantial decrease in viral load, whereas faster decay leads to more rapid viral resurgence. The parameters used in both simulations are $s = 100000$, $\lambda = 1 \times 10^{-6}$, $d = 0.1$, $h = 0.1$, $e = 1 \times 10^{-10}$, $d_1 = 1.0$, $p_1 = 10^5$, $c_1 = 29$, $A_0 = 0.3$, and $T_0 = 10^5$.

To keep  physiological accuracy, in this  work  we use  parameter values derived from experimental and clinical studies, as shown in the table below.
\begin{table}[h]
\centering
\caption{Physiological Parameters Used in the Model}
\renewcommand{\arraystretch}{1.2}
\resizebox{.95\columnwidth}{!}{ % Scale to fit within a column
\begin{tabular}{|c|c|c|}
\hline
\textbf{Parameter} & \textbf{Symbol} & \textbf{Value Range} \\ 
\hline
Latent cell reactivation rate & $\lambda$ & $0.01 - 0.1$ per day \\
Latent cell formation rate & $\xi$ & Variable \\
Latent cell decay rate & $\delta_L$ & $10^{-4} - 10^{-2}$ per day \\
Infected cell death rate & $\delta_I$ & $0.5 - 1.0$ per day \\
Target cell regeneration rate & $s$ & $10^3 - 10^5$ cells/µL/day \\
Target cell decay rate & $d$ & $10^{-2} - 10^{-1}$ per day \\
Viral infection rate & $k$ & $10^{-8} - 10^{-6}$ mL/virion/day \\
Virus production rate & $p$ & $10^2 - 10^4$ virions/cell/day \\
Virus clearance rate & $c$ & $10 - 30$ per day \\
Immune stimulation rate & $\alpha$ & $10^{-2} - 10^{-1}$ per day \\
Effector cell decay rate & $\mu$ & $10^{-2} - 10^{-1}$ per day \\
\hline
\end{tabular}
}
\label{tab:parameters}
\end{table}
This parameter table reflects the established physiological ranges while incorporating variability to account for patient heterogeneity.

Our analysis shows that the initial size of the latent reservoir significantly  affects infection dynamics. Our results depict that larger values of \( L_0 \) lead to a slower decline in  latent cells \( L(t) \)  which  also indicates  the extended time required for reservoir clearance. This delay results in elevated levels of infected cells \( I(t) \)  since more latent cells become activated over time. The viral load \( V(t) \) closely follows the trajectory of \( I(t) \).  As a result,  larger reservoirs sustain prolonged viral replication and delayed  clearance.

Our results  underscore the direct correlation between reservoir size, infection burden, and rebound duration. The  results of this paper   also  indicate  the importance of reducing the latent reservoir to shorten rebound phases and  also   to limit peak viral loads, which reinforces  the challenges of viral eradication. It also supports therapeutic strategies aimed at minimizing reservoir size and optimizing reactivation timing to enhance post-treatment control.

\section{Optimized Shock-and-Kill Strategy with ART Cycling and Latency Reversal Agents}

As discussed before,  the ``Shock-and-Kill'' strategy aims to eliminate latent HIV by using latency-reversing agents (LRAs) to activate dormant viruses, followed by clearance via ART or the immune system. We explore two approaches: (i) \textit{simultaneous ART and shock}, where LRAs and ART are applied together, and (ii) \textit{opposite-phase ART and shock}, where LRAs are administered during ART interruption. Simultaneous intervention reduces  rebound by blocking reinfection, but  this may  also reduce immune clearance  because of  limited antigen presentation and potential interference with LRA efficacy. In the country, the opposite-phase approach enhances immune recognition but risks uncontrolled replication, immune escape, and reservoir reseeding if clearance is delayed.

In this case the latent cell dynamics is governed by 
\begin{equation}
\frac{dL}{dt} = \xi V (1 - A) - \lambda (1 - A) L - \gamma S L - \delta_L L,
\end{equation}
where the terms represent, respectively: latent cell formation from viral exposure (\( \xi V (1 - A) \)), spontaneous reactivation suppressed by ART (\( \lambda (1 - A) L \)), LRA-induced reactivation (\( \gamma S L \)), and natural decay of latent cells (\( \delta_L L \)).

The infected cell population  is governed by  
\begin{equation}
\frac{dI}{dt} = \lambda (1 - A) L + \gamma S L + \beta V (1 - A) - \delta_I I - k A I.
\end{equation}
In this case,  infected cells arise from reactivated latency and new infections, while being cleared by natural decay and ART-mediated killing.

The viral load  is given by 
\begin{equation}
\frac{dV}{dt} = p I (1 - A) - c V - \epsilon A V. 
\end{equation}
Here \(A(t) = A_0 e^{-K_{drug} t}\) and \(S(t) = S_0 e^{-k_S t}\) represent the pharmacokinetic decay of ART and LRA with respective rates \(K_{drug}\) and \(k_S\).

\begin{figure}[ht]
    \centering
    \includegraphics[width=0.8\linewidth]{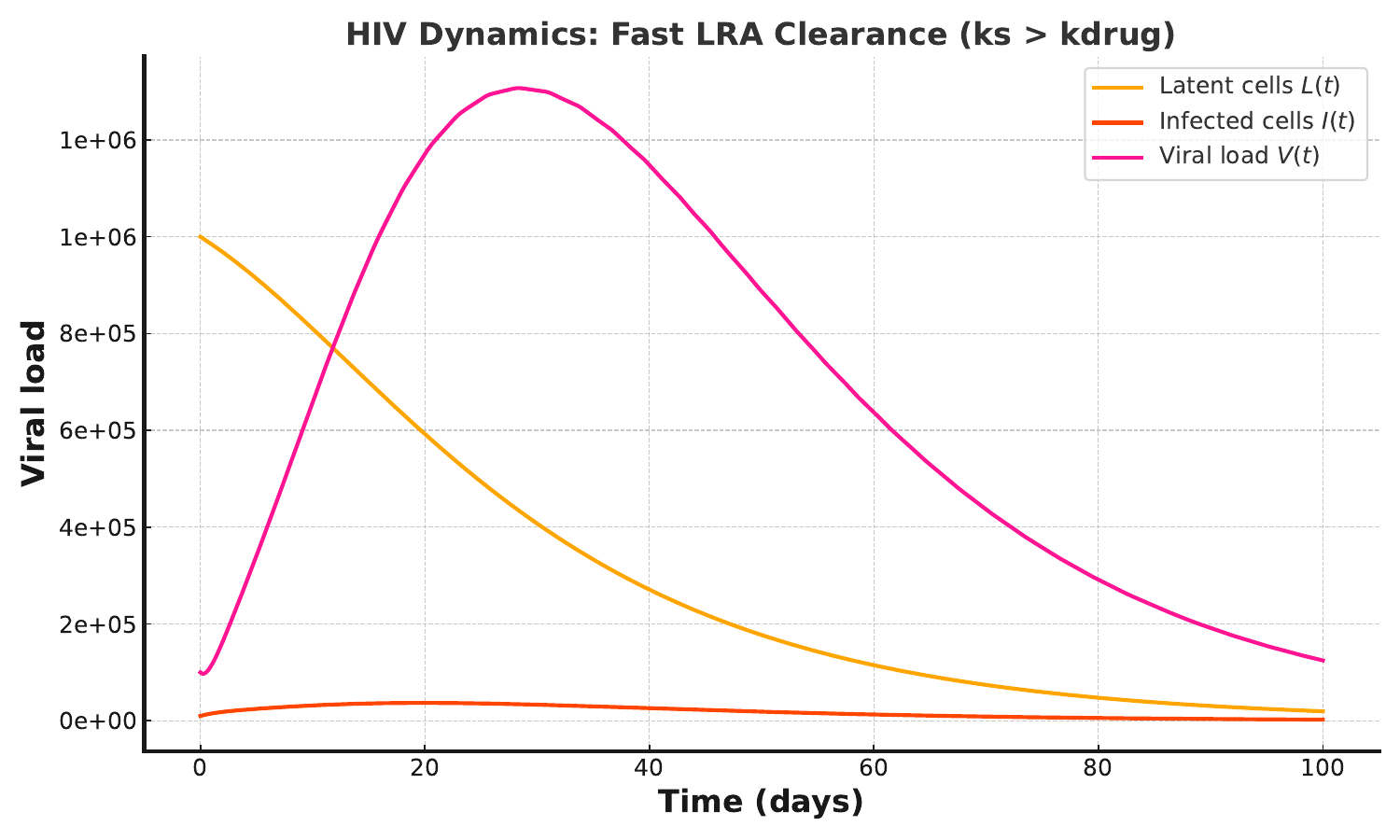}
    \caption{Figure shows the time  evolution of HIV dynamics during  ART and LRA treatment. The plot depicts the dynamics of latent cells ($L$), infected cells ($I$), and viral load ($V$)  as  a  function of  time. The figure shows that ART ($A=0.8$) decreases  viral replication and infected cells, whereas LRA ($S=0.2$) induces reactivation of latent cells. The viral load ($V$) exhibits an initial decline because of  ART, but  it also  shows transient increases due to LRA-induced reactivation. We fix the  following parameters as  $V_0 = 10^5$, $L_0 = 10^6$, $I_0 = 10^4$, $\lambda = 0.05$, $\gamma = 0.02$, $\delta_L = 0.001$, $\delta_I = 0.5$, $\beta = 0.001$, $p = 50$, $c = 1$, $\epsilon = 0.3$, $k = 0.3$, $K_{drug}=0.05$  and $k_s=0.2$.}
    \label{fig:hiv_dynamics}
\end{figure}
\begin{figure}[ht]
    \centering
    \includegraphics[width=0.8\linewidth]{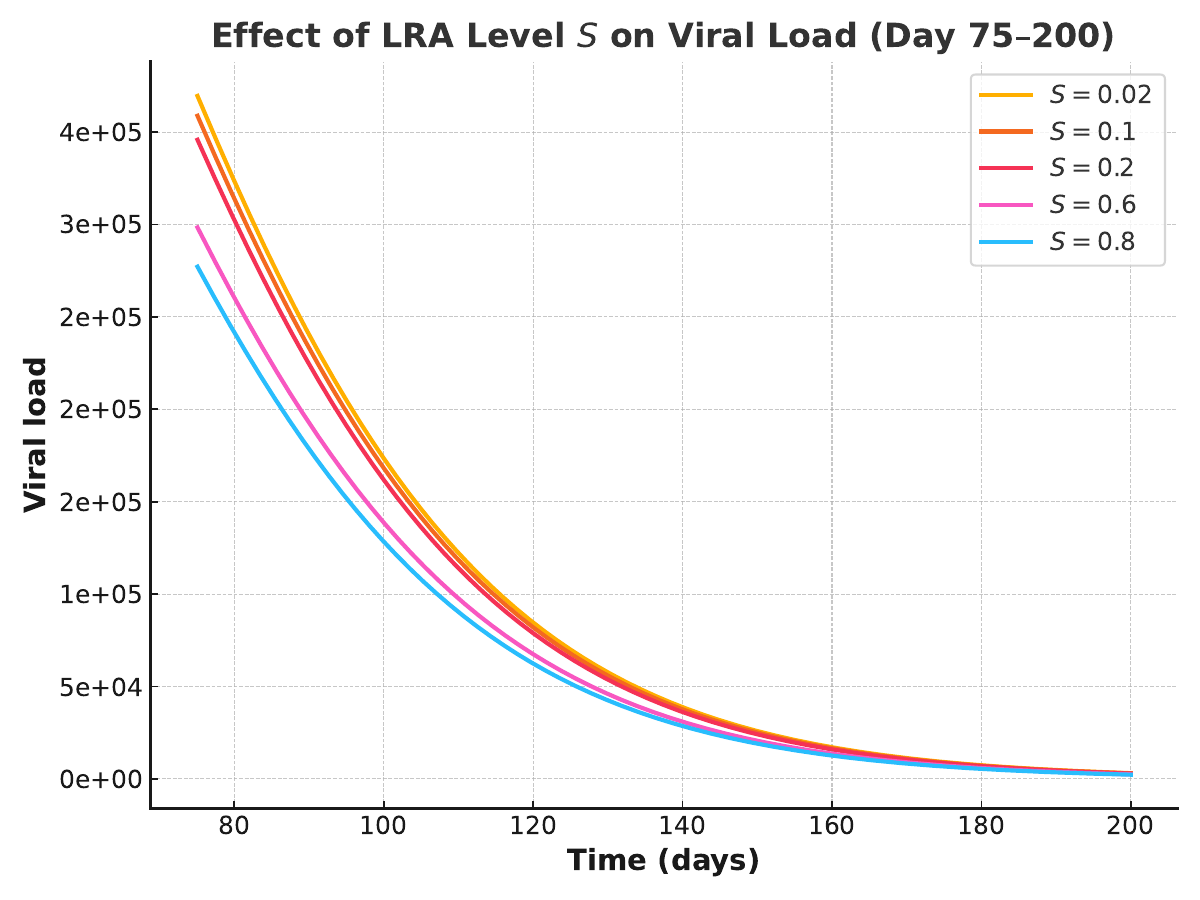}
    \caption{Effect of different LRA levels ($S$)  as a function of time   and  viral load ($V$). An increase in  $S$ leads to more latent cell activation, and this leads to  a transient increase in viral load. If $S$ is too high, the viral load rebounds significantly before the ART can be suppressed. We fix the parameters as  $V_0 = 10^5$, $L_0 = 10^6$, $I_0 = 10^4$, $\lambda = 0.05$, $\gamma = 0.02$, $\delta_L = 0.001$, $\delta_I = 0.5$, $\beta = 0.001$, $p = 50$, $c = 1$, $\epsilon = 0.3$, $k = 0.3$,  $K_{drug}=0.02$  and $k_s=0.05$.}
    \label{fig:effect_of_S_on_V}
\end{figure}
\begin{figure}[ht]
    \centering
    \includegraphics[width=0.8\linewidth]{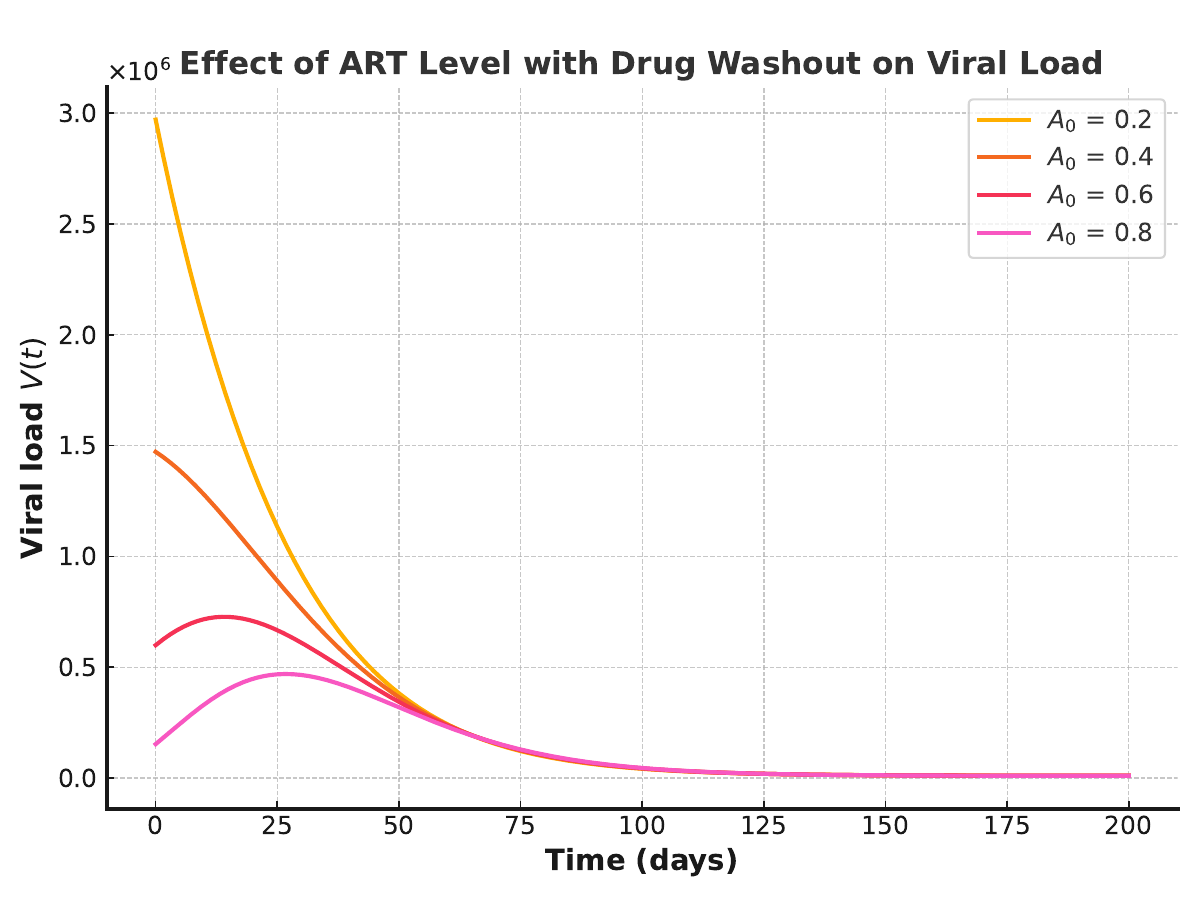}
    \caption{Effect of different ART levels ($A$) on viral load ($V$) over time. The figure depicts that a higher ART decreases  the viral load more effectively by stepping down the risk of viral rebound. If ART is too low, the viral load remains elevated, increasing the risk of reservoir reseeding. In the figure, we  fix  $V_0 = 10^5$, $L_0 = 10^6$, $I_0 = 10^4$, $\lambda = 0.05$, $\gamma = 0.02$, $\delta_L = 0.001$, $\delta_I = 0.5$, $\beta = 0.001$, $p = 50$, $c = 1$, $\epsilon = 0.3$, $k = 0.3$,  $K_{drug}=0.02$  and $k_s=0.05$.}
    \label{fig:effect_of_A_on_V}
\end{figure}
\begin{figure}[ht]
    \centering
    \includegraphics[width=0.8\linewidth]{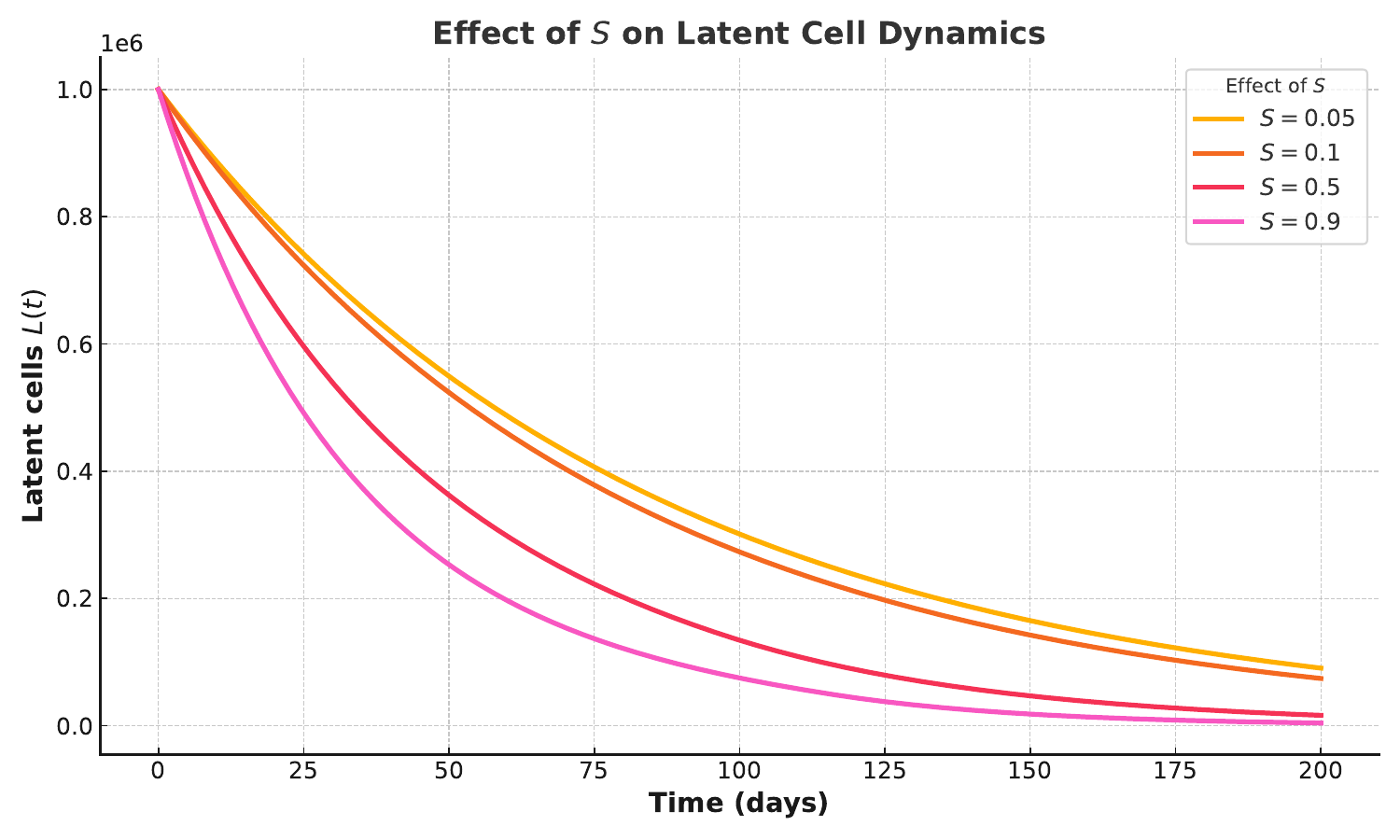}
    \caption{ LRA levels ($S$) as a function of time for different  latent cells ($L$)  reservoir. When  $S$ increases, it induces more latent cell activation, which  leads to a reduction in the latent reservoir. If $S$ is too high, a significant number of latent cells transition to active infected cells, which increases  viral replication.  We plot the figure by fixing $V_0 = 10^5$, $L_0 = 10^6$, $I_0 = 10^4$, $\lambda = 0.05$, $\gamma = 0.02$, $\delta_L = 0.001$, $\delta_I = 0.5$, $\beta = 0.001$, $p = 50$, $c = 1$, $\epsilon = 0.3$, $k = 0.3$  $K_{drug}=0.02$  and $k_s=0.05$.}
    \label{fig:effect_of_S_on_L}
\end{figure}

To examine how latent cells, viral loads, and infected cells respond to treatment parameters, we conducted numerical simulations of the deterministic model. Figure~9 depicts the time evolution of HIV dynamics under combined ART and LRA interventions. In this simulation, ART and LRA  are fixed at   $A = 0.8$ and LRA with $S = 0.2$. As shown in the figure, ART effectively suppresses viral replication and reduces the population of infected cells, at the same time  the LRA transiently reactivates latent cells. This leads  to a temporary increase in the viral load. Over time, the  latent cells, infected cells, and viral load decline to very low levels. This behavior is particularly pronounced  when the LRA decay rate exceeds that of ART ($k_s > k_{\mathrm{drug}}$), and this highlights  a regime where shock-and-kill is highly effective. These results are consistent with the stochastic analysis presented in SectionII, where activation clustering and delayed rebound are favored by slow ART washout and rapid LRA clearance. 

In Figure~10, we explore  how  different  levels of LRA stimulation ($S$) dictate the viral load ($V$) under the shock-and-kill framework. The figure depicts that higher values of $S$ enhance the activation of latent cells and this  leads  to an initial rise in viral load due to reactivation events. However, this transient increase latter  followed  by a steady decline in $V$  when  activated cells become susceptible to immune or ART-mediated clearance. Notably, when the LRA decay rate exceeds that of ART ($k_s > k_{\mathrm{drug}}$), the viral load is effectively reduced across time. This reflects a favorable therapeutic window in which activation is strong enough to expose the latent reservoir but short-lived enough to avoid uncontrolled rebound. As $S$ increases, the clearance of infected cells is accelerated.  This, in turn, ultimately results in a net suppression of viral replication. Our  findings support the strategic use of potent but transient LRA stimulation combined   with  ART, which decays  slowly  to maximize the efficacy of shock-and-kill interventions.

  The effect of ART levels ($A$) on viral dynamics is depicted  in Figure~11. As shown in the figure, higher ART concentrations lead to a pronounced suppression of the viral load ($V$) and this reduces  the likelihood of rebound and reservoir reseeding. On the other hand, suboptimal ART levels fail to fully contain replication, which allows  persistent infection and indicates the necessity of sustained ART coverage during reactivation. 

In Figure~12, we examine the time evolution of LRA stimulation ($S$) and its impact on the latent reservoir ($L$). The simulation results  show that an increase  in  $S$ leads to faster and more extensive activation of latent cells and accelerated reservoir depletion. This effect is especially pronounced when the LRA decay rate exceeds that of ART ($k_s > k_{\mathrm{drug}}$) and this creates a temporal window in which shock is effective and clearance is unhindered. These results underscore the importance of balancing LRA potency and pharmacokinetics in order to maximize latent cell clearance while minimizing rebound risk.

At equilibrium, after complete drug washout (\( A \to 0 \), \( S \to 0 \)), the steady-state expressions are
\begin{equation}
L^* = \frac{\xi V^*}{\lambda + \delta_L},
\end{equation}
\begin{equation}
I^* = \frac{\lambda L^* + \beta V^*}{\delta_I},
\end{equation}
\begin{equation}
V^* = \frac{p I^*}{c}.
\end{equation}

It is evident   that the interplay between ART and LRA dynamics critically dictates  HIV reactivation and reservoir depletion. The results discussed in this work  indicate that  high ART levels are essential for sustained viral suppression  at the same time  well-timed and appropriately dosed LRAs can effectively induce latent cell reactivation. However, excessive LRA stimulation without adequate ART coverage risks viral rebound and reservoir reseeding. Moreover, the size of the latent reservoir directly influences the rebound dynamics. As a result, larger reservoirs result in  prolonged infection and delayed clearance.

These findings stress  the importance of synchronizing ART and LRA timing  in order to  maximize shock-and-kill efficacy. An optimal balance (at which  ART effectively suppresses replication and LRAs strategically expose latent cells) offers a promising strategy for decreasing  the reservoir and minimizing post-treatment rebound. Thus, we believe  that this work provides a robust theoretical framework to inform the design of time-sensitive, pharmacokinetically aligned HIV cure strategies.

\section{Summary and Conclusion}

In this work, we  use  a stochastic framework to model HIV latency reversal that incorporates  time-dependent immune fluctuations, ART pharmacokinetics, and Gamma-distributed waiting times. Our results show that rebound timing is highly sensitive to ART decay. A slower clearance prolongs latency, while rapid decay accelerates reactivation. We analyzed four activation profiles—constant, sinusoidal, stochastic, and decaying—and found that immune-driven oscillations and stochastic fluctuations shape reactivation dynamics, with strong noise increasing rebound risk.

The results obtained  in this study  have direct implications for \textit{ shock-and-kill } strategies. Our results  depict  that synchronizing LRA-induced activation with immune peaks enhances viral clearance, while slower ART decay extends the therapeutic window. Overall, optimizing the timing of interventions and addressing stochastic variability are critical for reducing reservoirs and achieving durable post-treatment control.

In conclusion,  our results  stress  the need for the right timing in HIV eradication strategies. Aligning latency-reversing interventions with periods of heightened immune activity could  not only enhance treatment efficacy  but also  improve post-treatment viral control. It  should be noted that addressing the stochastic nature of reactivation remains a key challenge since this  requires  a combination of mathematical modeling, immunological insights, and precise  medicine approaches to optimize long-term remission strategies.

\section*{Acknowledgment}
I would like to thank  Mulu  Zebene and Asfaw Taye for the
constant encouragement. 

\section*{Data Availability Statement }This manuscript has no
associated data or the data will not be deposited. [Authors’
comment: Since we presented an analytical work, we did not
collect any data from simulations or experimental observations.]


\section*{References}

\begin{thebibliography}{120}

\bibitem{Murray2016} Murray AJ, Kwon KJ, Farber DL, Siliciano RF. The latent reservoir for HIV-1: How immunologic memory and clonal expansion contribute to HIV-1 persistence. 	extit{J Immunol}. 2016;197(2):407–417.
\bibitem{Whitney2014} Whitney JB, et al. Rapid seeding of the viral reservoir prior to SIV viraemia in rhesus monkeys. 	extit{Nature}. 2014;512(7512):74–77.
\bibitem{Okoye2018} Okoye AA, Picker LJ. CD4+ T-cell depletion in HIV infection: mechanisms of immunological failure. 	extit{Immunol Rev}. 2018;254(1):54–64.
\bibitem{Chun1997} Chun TW, et al. Presence of an inducible HIV-1 latent reservoir during highly active antiretroviral therapy. 	extit{Proc Natl Acad Sci U S A}. 1997;94(24):13193–13197.
\bibitem{Siliciano2003} Siliciano RF, et al. Long-term follow-up studies confirm the stability of the latent reservoir for HIV-1 in resting $CD4+$ T cells. 	extit{Nat Med}. 2003;9(6):727–728.
\bibitem{Li2020} Li B, et al. Proviruses with identical sequences comprise a large fraction of the replication-competent HIV reservoir. 	extit{Proc Natl Acad Sci U S A}. 2020;117(8):3886–3893.



\bibitem{Wu2020} Wu G, et al. The landscape of HIV-1 transcription in vivo. 	extit{Nat Microbiol}. 2020;5(3):438–451.
\bibitem{Pinkevych2019} Pinkevych M, et al. Modeling of experimental data supports HIV reactivation from latency after treatment interruption at high but variable rates. 	extit{Elife}. 2019;8:e49022.
\bibitem{Fennessey2017} Fennessey CM, Lifson JD. Immune control of HIV-associated viral reservoirs. 	extit{Curr Opin HIV AIDS}. 2017;12(2):150–155.
\bibitem{Byrareddy2016} Byrareddy SN, et al. Sustained virologic control in $SIV+$ macaques after antiretroviral and $\alpha 4 \beta 7$ antibody therapy. 	extit{Science}. 2016;354(6309):197–202.
\bibitem{Whitney2018} Whitney JB, et al. Combination anti-HIV antibodies provide sustained virological suppression. 	extit{Nature}. 2018;561(7724):479–484.
\bibitem{Hill2018} Hill AL, et al. Modeling HIV-1 reactivation and clearance in resting $CD4+$ T cells. 	extit{PLoS Pathog}. 2018;14(11):e1007333.
\bibitem{Hill2014} Hill AL, Rosenbloom MT, Siliciano RF, Mark DS. Insufficient evidence for rare activation of latent HIV in the absence of stochastic effects. 	extit{PLoS Pathog}. 2014;10(2):e1004000.
\bibitem{mes2}
S. Wang, Y. Pan, Q. Wang, H. Miao, A. N. Brown and L. Rong,  Mathematical Biosciences 
  {\bf 328}, 108438 (2020).
\bibitem{mes1}
C. Zitzmann and L. Kaderali, Forntiers in Microbiology  {\bf 9}, 1546 (2018).
\bibitem{mes3}
Y. Wang, J. Liu and L. Liu, Advances in Difference Equations {\bf 1}, 225 (2018).
\bibitem{mes4}
S. S. Chen, C. Y. Cheng and Y. Takeuchi, Journal of Mathematical Analysis and Applications {\bf 442}, 642 (2016).
\bibitem{mu1}
A. S. Perelson,  D. E. Kirschner and R. De Boer , Math. Biosci  {\bf 114}, 81 (1993).
\bibitem{mu2}
A. S. Perelson, , A. U. Neumann, M. Markowitz, J. M Leonard  and D. D Ho  , Science   {\bf 271}, 1582 (1996).
\bibitem{mu3}
A. S. Perelson,  P. Essunger,  Y. Cao,  M. Vesanen, A. Hurley and K. Saksela, Nature {\bf 387}, 188  (1997).
\bibitem{mu4} Ho, D. D., Neumann, A. U., Perelson, A. S., Chen, W., Leonard, J. M., and
Markowitz,M. (1995). Rapid turnover of plasma virions and CD4 lymphocytes
in HIV-1 infection. Nature 373, 123–126. doi: 10.1038/373123a.
\bibitem {mu5} Bonhoeffer, S., May, R. M., Shaw, G. M., and Nowak, M. A. (1997). Virus
dynamics and drug therapy. Proc. Natl. Acad. Sci. U.S.A. 94, 6971–6976.
doi: 10.1073/pnas.94.13.6971.
\bibitem {mu6} Stafford, M. A., Corey, L., Cao, Y., Daar, E. S., Ho, D. D., and Perelson, A. S. (2000). Modeling plasma virus concentration during primary HIV infection. J. Theor.
Biol. 203, 285–301. doi: 10.1006/jtbi.2000.1076
\bibitem {mu7}Wei, X., Ghosh, S. K., Taylor, M. E., Johnson, V. A., Emini, E. A., Deutsch, P.,
et al. (1995).Viral dynamics in human immunodeficiency virus type 1 infection.
Nature 373, 117–122. doi: 10.1038/373117a0
\bibitem {mu8}Neumann, A. U. (1998). Hepatitis C viral dynamics in vivo and the antiviral efficacy of interferon- therapy. Science 282, 103–107. doi: 10.1126/science.282.5386.103
\bibitem {mu9}Dahari, H., Lo, A., Ribeiro, R. M., and Perelson, A. S. (2007a). Modeling hepatitis C virus dynamics: Liver regeneration and critical drug efficacy. J. Theor. Biol.
247, 371–381. doi: 10.1016/j.jtbi.2007.03.006
\bibitem {mu10}Nowak, M. A., and Bangham, C. R. (1996). Population dynamics of immune
responses to persistent viruses. Science 272, 74–79.
\bibitem {mu11}Nowak,M. A., Bonhoeffer, S., Hill, A.M., Boehme, R., Thomas, H. C., andMcDade,
H. (1996). Viral dynamics in hepatitis B virus infection. Proc. Natl. Acad. Sci.
U.S.A. 93, 4398–4402. doi: 10.1073/pnas.93.9.4398
\bibitem {mu12}Perelson, A. S. (2002). Modelling viral and immune system dynamics. Nat. Rev.
Immunol. 2, 28–36. doi: 10.1038/nri700
\bibitem {mu14}Wodarz, D., and Nowak,M. A. (2002).Mathematical models of HIV pathogenesis
and treatment. BioEssays 24, 1178–1187. doi: 10.1002/bies.10196
\bibitem{Pinkevych2016} Pinkevych M, et al. Frequency of HIV rebound after stopping antiretroviral therapy: A Markov model analysis. 	extit{PLoS Pathog}. 2016;12(8):e1005789.
\bibitem{DeScheerder2019} De Scheerder M, et al. HIV rebound is predominantly fueled by genetically identical viral expansions from diverse reservoirs. 	extit{Cell Host Microbe}. 2019;26(3):347–358.e7.
\bibitem{Lorenzo2016} Lorenzo-Redondo C, et al. Persistent HIV-1 replication maintains the tissue reservoir during therapy. 	extit{Nature}. 2016;530(7588):51–56.
\bibitem{Christensen-Quick2018} Christensen-Quick A, et al. Suboptimal CD8+ T cell-mediated immune surveillance in HIV-1 infection. 	extit{J Clin Invest}. 2018;128(5):2076–2089.
\bibitem{Zens2014} Zens KD, Farber DL. Memory CD4 T cells in influenza. 	extit{Curr Opin Virol}. 2014;9:102–108.
\bibitem{Bukrinsky1991} Bukrinsky M. The nuclear import of HIV-1 preintegration complexes. 	extit{J Virol}. 1991;65(2):638–645.
\bibitem{Ribeiro2022} Ribeiro RM, Perelson AS. Modeling HIV persistence during antiretroviral therapy. 	extit{Curr Opin Virol}. 2022;53:101216.

\bibitem{hbv1} Li X, Li M, Tian Y, Liu R, Khan SU. A stochastic hepatitis B model with media coverage and Lévy noise. \textit{J Taibah Univ Sci}. 2024;18(1):2414523. doi:10.1080/16583655.2024.2414523.
\bibitem{hbv2}Shah SMA, Nie Y, Din A, Alkhazzan A, Younas B. Stochastic modeling and analysis of hepatitis and tuberculosis co-infection dynamics. \textit{Chin Phys B}. 2024;33(11):110203. doi:10.1088/1674-1056/ad7afa.
\bibitem{hbv3} Shah SMA, Nie Y, Din A, Alkhazzan A. Dynamics of hepatitis B virus transmission with a Lévy process and vaccination effects. \textit{Mathematics}. 2024;12(11):1645. doi:10.3390/math12111645.
\bibitem{hbv4} Shah SMA, Nie Y, Din A, Alkhazzan A. Stochastic optimal control analysis for HBV epidemic model with vaccination. \textit{Symmetry}. 2024;16(10):1306. doi:10.3390/sym16101306.
\bibitem{hbv5} Shah SMA, Nie Y, Din A, Alkhazzan A, Arshad A, Younas B. Stochastic modeling for the transmission of hepatitis B virus with multiple time-delays and vaccination effect. \textit{Arab J Math}. 2025;2025:Article 00534. doi:10.1007/s40065-025-00534-y.
\bibitem{worm1} Shah SMA, Tahir H, Khan A, Khan WA, Arshad A. Stochastic model on the transmission of worms in wireless sensor networks. \textit{J Math Tech Modeling}. 2024;1(1):31. doi:10.56868/jmtm.v1i1.31.
\bibitem{worm2} Alkhazzan A, Wang J, Nie Y, Shah SMA, Almutairi DK, Khan H, Alzabut J. Lyapunov‑based analysis and worm extinction in wireless networks using stochastic SVEIR model. \textit{Alexandria Eng J}. 2025;118:337–353. doi:10.1016/j.aej.2025.01.040.
\bibitem{burke2022}
Alkhazzan A, Loqman IGH, Shah SMA, Alkhazzan A. Existence, uniqueness, and stability analysis of the fractional-order Burke–Shaw model with ABC‑fractional derivative. \textit{Sana’a Univ J Appl Sci Technol}. 2025;3(2):690–697. doi:10.59628/jast.v3i2.1470.





\bibitem{Davenport2019} Davenport MP. Mathematical modeling of HIV latency and reactivation. 	extit{Trends Microbiol}. 2019;27(9):750–759.
\bibitem{Lewis2015} Lewis GK, Pazgier M. Experimental strategies for HIV latency reversal. 	extit{Annu Rev Virol}. 2015;2(1):607–628.
\bibitem{Bosque2009} Bosque A, Planelles V. Induction of HIV-1 latency and reactivation in primary memory CD4+ T cells. 	extit{Blood}. 2009;113(1):58–65.
\bibitem{Marsden2018} Marsden MD, et al. Latency reversal and viral clearance to cure HIV-1. 	extit{Trends Microbiol}. 2018;26(10):833–848.
\bibitem{Borducchi2018} Borducchi EN, et al. Antibody and TLR7 agonist delay viral rebound in SHIV-infected monkeys. 	extit{Nature}. 2018;563(7731):360–364.
\bibitem{Archin2012} Archin NM, et al. Administration of vorinostat disrupts HIV-1 latency in patients on antiretroviral therapy. 	extit{Nature}. 2012;487(7408):482–485.
\bibitem{van} Van Dorp CH, Conway JM, Barouch DH, Whitney JB, Perelson AS. Models of SIV rebound after treatment interruption that involve multiple reactivation events. 	extit{PLoS Comput Biol}. 2020;16(10):e1008241.
\bibitem{pink} Wu Y, Pinkevych M, Xu Z, Keele BF, Davenport MP, Cromer D. Impact of fluctuation in frequency of human immunodeficiency virus/simian immunodeficiency virus reactivation during antiretroviral therapy interruption. 	extit{Proc R Soc B}. 2020;287(1930):20200354.
\bibitem{yu2023}
Yu X, Zhao L, Yuan Z, Li Y. Pharmacokinetic drug-drug interactions involving antiretroviral agents: An update. \textit{Curr Drug Metab}. 2023;24(7):493--524. doi:10.2174/1389200224666230418093139.






\end{thebibliography}
\end{document}